\documentclass[aps,prb,reprint,groupedaddress,superscriptaddress,twocolumn]{revtex4-2}

\setcitestyle{super}
\usepackage[pdftex]{graphicx}
\usepackage{dcolumn}
\usepackage{bm}
\usepackage{amsmath}
\usepackage{latexsym}
\usepackage{epstopdf}
\usepackage{color}
\usepackage{lipsum}
\usepackage{xargs}

\definecolor{bluenp}{rgb}{0, 0.412, 0.647}
\usepackage{hyperref}
\hypersetup{colorlinks=true, linkcolor=bluenp, citecolor=bluenp, urlcolor=bluenp,anchorcolor=bluenp}

\newcommand{\rfig}[1]{Fig.~\ref{#1}}

\newcommand{\meth}[1]{\hyperref[#1]{\textcolor{black}{Methods}}}

\definecolor{brickred}{rgb}{0.79, 0.25, 0.33}

\definecolor{green}{rgb}{0.1, 0.9, 0.1}

\newenvironment{sequation}{\begin{equation}\small}{\end{equation}}
\newenvironment{seqnarray}{\begin{eqnarray}\small}{\end{eqnarray}}
\def\rm{\mathrm}

\DeclareMathSizes{10}{9}{7}{5}

\newcommand{\newsection}[1]{
    \vspace{12pt}
    \noindent
    \begingroup
    \fontsize{12pt}{12pt}\selectfont
    \textbf{#1}
    \endgroup
    \par\noindent
}

\newcommand{\mysection}[1]{
    \vspace{6pt}
    \noindent
    \begingroup
    \fontsize{11pt}{12pt}\selectfont
    \textbf{#1}
    \endgroup
    \par\noindent
}

\newcommand{\mysubsection}[1]{
    \vspace{6pt}
    \noindent
    \fontsize{10pt}{12pt}\selectfont
    \textbf{#1}
    \par\noindent
}

\makeindex

\begin{document}
\title{Distributed multi-parameter  quantum metrology with a superconducting quantum network}
\author{Jiajian Zhang}
    \thanks{These authors contributed equally to this work.}
	\affiliation{International Quantum Academy, Shenzhen 518048, China}

    \author{Lingna Wang}
    \thanks{These authors contributed equally to this work.}
    \affiliation{Department of Mechanical and Automation Engineering, The Chinese University of Hong Kong, Shatin, Hong Kong}
    
    \author{Yong-Ju Hai}
    \thanks{These authors contributed equally to this work.}
	\affiliation{International Quantum Academy, Shenzhen 518048, China}

	\author{Jiawei Zhang}
	\thanks{These authors contributed equally to this work.}
 	\affiliation{Southern University of Science and Technology, Shenzhen 518055, China}
  	\affiliation{International Quantum Academy, Shenzhen 518048, China}
	
	\author{Ji Chu}
	\affiliation{International Quantum Academy, Shenzhen 518048, China}

    \author{Ji Jiang}
	\affiliation{International Quantum Academy, Shenzhen 518048, China}

	\author{Wenhui Huang}
	\affiliation{International Quantum Academy, Shenzhen 518048, China}
	
	\author{Yongqi Liang}
 	\affiliation{Southern University of Science and Technology, Shenzhen 518055, China}
  	\affiliation{International Quantum Academy, Shenzhen 518048, China}

	\author{Jiawei Qiu}
	\affiliation{International Quantum Academy, Shenzhen 518048, China}
	
	\author{Xuandong Sun}
 	\affiliation{Southern University of Science and Technology, Shenzhen 518055, China}
  	\affiliation{International Quantum Academy, Shenzhen 518048, China}

	\author{Ziyu Tao}
	\affiliation{International Quantum Academy, Shenzhen 518048, China}
	
	\author{Libo Zhang}
 	\affiliation{Southern University of Science and Technology, Shenzhen 518055, China}
  	\affiliation{International Quantum Academy, Shenzhen 518048, China}

	\author{Yuxuan Zhou}
	\affiliation{International Quantum Academy, Shenzhen 518048, China}

	\author{Yuanzhen Chen}
 	\affiliation{Southern University of Science and Technology, Shenzhen 518055, China}

    \author{Weijie Guo}
	\affiliation{International Quantum Academy, Shenzhen 518048, China}
	
	\author{Xiayu Linpeng}
	\affiliation{International Quantum Academy, Shenzhen 518048, China}

	\author{Song Liu}
	\affiliation{International Quantum Academy, Shenzhen 518048, China}
    \affiliation{Shenzhen Branch, Hefei National Laboratory, Shenzhen 518048, China}

	\author{Wenhui Ren}
	\affiliation{International Quantum Academy, Shenzhen 518048, China}

    \author{Youpeng Zhong}
    \email{zhongyoupeng@iqasz.cn}
    \affiliation{International Quantum Academy, Shenzhen 518048, China}
    \affiliation{Shenzhen Branch, Hefei National Laboratory, Shenzhen 518048, China}
 
    \author{Jingjing Niu}
	\email{niujj@iqasz.cn}
	\affiliation{International Quantum Academy, Shenzhen 518048, China}
 \affiliation{Shenzhen Branch, Hefei National Laboratory, Shenzhen 518048, China}
    
    \author{Haidong Yuan}
    \email{hdyuan@mae.cuhk.edu.hk}
    \affiliation{Department of Mechanical and Automation Engineering, The Chinese University of Hong Kong, Shatin, Hong Kong}
    
    \author{Dapeng Yu}
    \affiliation{International Quantum Academy, Shenzhen 518048, China}
    \affiliation{Shenzhen Branch, Hefei National Laboratory, Shenzhen 518048, China}

\maketitle

\noindent
{\bf{Quantum metrology has emerged as a powerful tool for timekeeping, field sensing, and precision measurements in fundamental physics. With the advent of distributed quantum metrology, its capabilities have extended to probing spatially distributed parameters across networked quantum systems. However, scalable implementations of distributed quantum metrology with multi-parameter estimation remain limited, particularly due to the challenges of generating and distributing entanglement across a quantum network and dealing with incompatibilities in multi-parameter quantum metrology. Here we demonstrate distributed multi-parameter quantum metrology on a modular superconducting quantum network with low-loss microwave interconnects, a platform that uniquely combines fast gate operations, adaptive control, and deterministic non-local entanglement generation. Using a control-enhanced sequential protocol, we estimate all three components of a remote vector field, achieving up to 13.72 dB improvement in precision over the individual strategy. We further perform direct estimation of vector field gradients along two directions across spatially separated nodes, realizing a 3.44 dB gain over local entanglement strategies. These results establish superconducting quantum networks as a competitive and reconfigurable platform for scalable multi-parameter distributed quantum metrology.}}

\bigskip


\noindent 
The quest for high-precision measurement is fundamental to scientific advancement. Quantum metrology, which exploits quantum resources such as superposition and entanglement, enables measurement precision beyond classical limits~\cite{giovannetti2011advances, toth2014quantum, RevModPhys.90.035005, PhysRevLett.96.010401, PhysRevLett.102.100401}. 
Distributed quantum metrology (DQM) extends this advantage to spatially separated quantum sensors, allowing the characterization of remote or distributed signals~\cite{proctor2018multiparameter,Ge2018,Sekatski2020,gessner2020multiparameter,Zhang2021,Oh2022,Zhuang_2020,Fadel_2023,PhysRevA.108.032621,PhysRevResearch.6.013246}. It has broad applications from network clock synchronization~\cite{Giovannetti2001,Komar2014} to gravitational and magnetic field mapping~\cite{Marra2018,schnabel2010quantum,RevModPhys.89.035002, taylor2008high}. 
Recent demonstrations using photonic~\cite{LiuBY2024,Liu2020,Xia2020,guo2020distributed,hong2021quantum,PhysRevX.11.031009,cimini2023deep,kim2024distributed} and atomic~\cite{Malia2022} platforms have proven that entanglement distributed across nodes can significantly enhance single-parameter sensing. 
These implementations typically measure global properties, such as an average phase, encoded by mutually commuting generators. Despite the progress, a critical open challenge for practical distributed quantum metrology is the extension to multi-parameter sensing with non-commuting generators. In this more complex scenario, conventional estimation strategies are highly inefficient~\cite{proctor2018multiparameter, PhysRevLett.116.030801, ALBARELLI2020126311, Fadel_2023, Pezze2025AdvMultipara}. Successfully overcoming this barrier will require two key advancements: first, the generation of high-quality, genuine non-local entanglement across the quantum network, and second, the design of metrological protocols capable of simultaneously estimating multiple parameters with high precision ~\cite{xia2023toward, chen2024simultaneous, Yuan2016, Hou2020, Hou2019, Hou2021}. 

\begin{figure*}[htbp]
\begin{center}
	\includegraphics[width=0.8\textwidth]{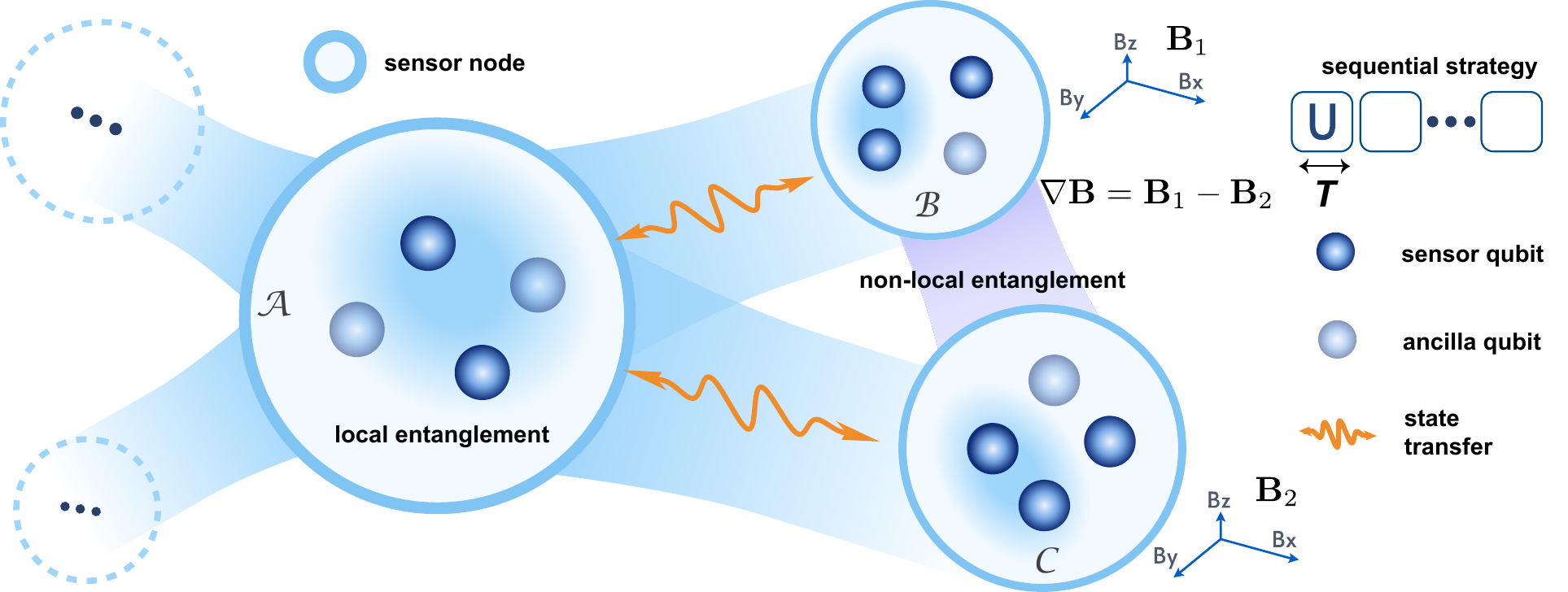}
	\caption{
		\label{fig1}
		{\bf Schematic of distributed quantum metrology with a modular superconducting sensor network.} 
        The platform comprises a central module $\mathcal{A}$ and multiple spatially separated sensor modules ($\mathcal{B}$, $\mathcal{C}$), each hosting several transmon qubits (blue and light blue spheres). The sensor modules are connected to the central module via low-loss coaxial cables, enabling high-fidelity microwave state transfer (orange arrows) and the generation of non-local entanglement across remote nodes. This architecture supports two key sensing protocols: (1) estimation of remote vector fields encoded by locally non-commuting generators (e.g., $\mathbf{B}_1$, $\mathbf{B}_2$), and (2) estimation of spatial gradients $\nabla \mathbf{B}$ via distributed entangled probes. Each sensor qubit undergoes a sequence of signal and control unitaries with interrogation time $T$ (top right), implementing a sequential metrological strategy. The blue background indicates local connectivity within a module, while the purple shading highlights modules linked by non-local entanglement. The inclusion of both sensor and ancilla qubits reflects their complementary roles in enabling tailored entangled state preparation, signal encoding, and joint measurement for multi-parameter sensing. }
\end{center}
\end{figure*}

Superconducting circuits offer a powerful platform for DQM, uniquely combining high-speed quantum operations with scalable networking capabilities. Their nanosecond-scale gate times and native compatibility with microwave control and signal transduction~\cite{Assouly2023,Sai2024,Deng2024,Wang2019SunL} make them particularly well-suited for detecting fast, weak signals, such as those arising in dark matter and cosmic-ray detections~\cite{Dixit2021,Tang2024,PRXQuantum.5.040323,li2025cosmic}. At the same time, advances in superconducting quantum networking have enabled high-fidelity quantum links~\cite{Kurpiers2018,Axline2018,CampagneIbarcq2018,Leung2019,Magnard2020,Zhong2021,Burkhart2021,CarreraVazquez2024} and deterministic entanglement distribution across modular architectures~\cite{Niu2023}, positioning this platform to implement advanced DQM protocols with real-time feedback and dynamic reconfigurability.

In this work, we demonstrate distributed multi-parameter quantum metrology in a modular superconducting quantum processor network~\cite{Niu2023}. Modular architectures provide a scalable path to quantum advantage by interconnecting specialized nodes—dedicated to  storage~\cite{Liu2024,Knaut2024,Zhang2022}, processing~\cite{Hoch2024,wei2025universal}, error correction~\cite{Zhou2018,Cai2024,PhysRevLett.129.240502}, or sensing~\cite{Komar2014,Mukhopadhyay2024}—via high-coherence quantum links~\cite{Niu2023}. Leveraging this framework, we demonstrate multi-parameter sensing of a remote 3D vector field using non-local entanglement between a central module and a sensor module, achieving a precision enhancement of up to 13.72 dB in variance over the individual measurement strategy. 
Furthermore, by creating a distributed 4-qubit GHZ state across two sensor modules via entanglement routing through the central node, we directly estimate the gradients of a spatially varying vector field along two different directions. This yields a 3.44 dB reduction in total variance over strategies with only local entanglement. These results underscore the potential of superconducting modular platforms for building scalable, high-speed, and reconfigurable quantum sensor networks.

\newsection{Results}

\mysubsection{Experimental setup}

\noindent
The experimental setup, as depicted in \rfig{fig1}, implements a modular superconducting quantum network in a star topology~\cite{Niu2023}. It comprises multiple sensor modules connected to a central module $\mathcal{A}$ via four 25-cm aluminum coaxial cables, which serve as low-loss transmission lines for microwave photons. Each module hosts four transmon qubits, and the interface between the qubit and cable is equipped with a tunable coupler that enables programmable interaction between qubits and cable modes. The cables function as multimode resonator buses that support standing wave modes, enabling the coherent transfer of microwave photons between modules~\cite{Niu2023}. By carefully coordinating controls on the qubits and couplers, the system can achieve high-fidelity inter-module operations, with state transfer efficiencies approaching 99\%. To generate and distribute entanglement between modules, entangled states are first prepared locally within the central module. The quantum state of one or more qubits is then coherently transferred to remote modules. We then leverage this non-local distributed entanglement to conduct two distinct metrological experiments.

\begin{figure*}[!htbp]
	\begin{center}
		\includegraphics[width=0.9\textwidth]{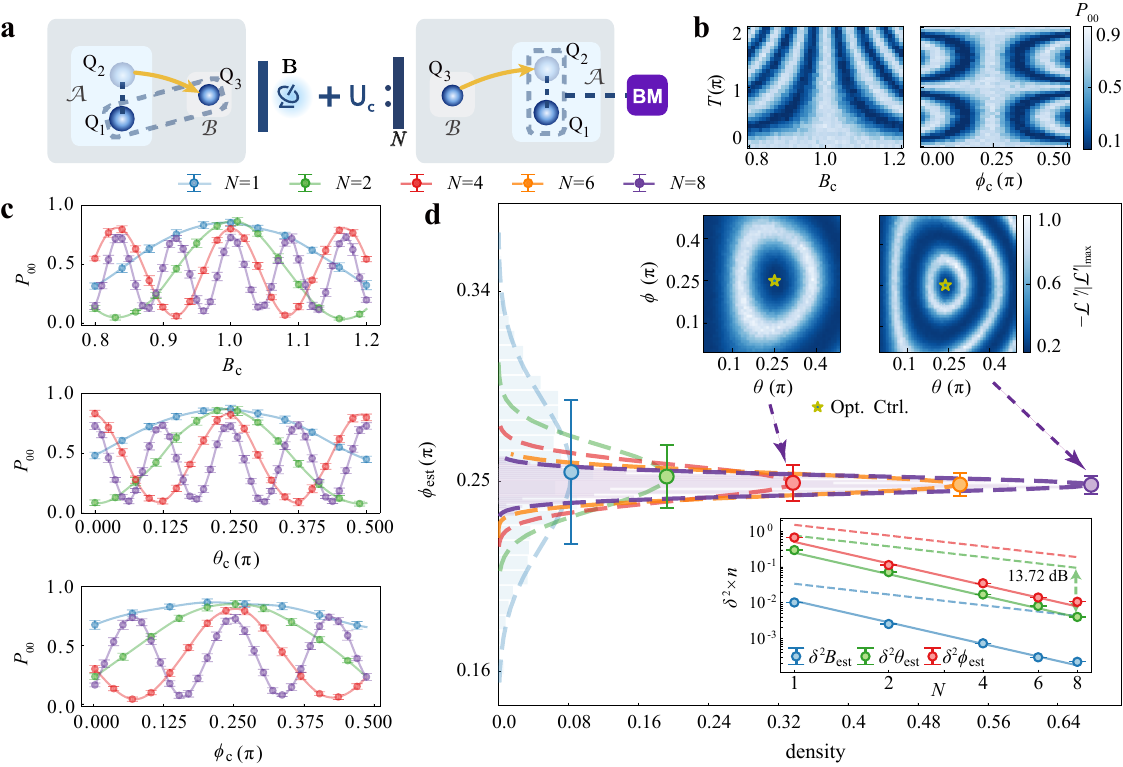}
		\caption{
			\label{fig2}
			{\bf Multi-parameter quantum metrology of a remote magnetic field with the sensor-ancilla network.} 
			{\bf a,} Schematic diagram of the controlled-enhanced sequential strategy. The protocol measures a local magnetic field at the sensor's position using a sequential strategy with one ancilla qubit. The three-component field $\mathbf{B}$ is parameterized by ${B, \theta, \phi}$. 
			{\bf b,} Measurement probability $P_{00}$ for $N=8$ cycles as a function of signal encoding time $T$ (from $0$ to $2\pi$), scanned across control parameters $B_c$ (left panel) and $\phi_c$ (right panel). 
			{\bf c,} Measurement probability $P_{00}$ for a fixed encoding time $T=1.5\pi$ at different cycle numbers $N=1,2,4,8$, scanned across control parameters ${ B_c, \theta_c, \phi_c }$. Control parameters not being scanned in {\bf b} and {\bf c} are fixed to their optimal values. Error bars denote the standard deviation. 
			{\bf d,} Results of parameter estimation. Main panel: Distribution of estimator $\phi_{\rm{est}}$ for $N=1,2,4,6,8$. Dashed lines represent Gaussian fitting of the estimator histogram; circles and the error bars mark the average value of $\phi_{\rm{est}}$ and the standard deviation. Upper inset: Landscape of the likelihood function at $N=4$ (left) and $N=8$ (right) for parameters $\phi$ and $\theta$. The star marks the optimal control setting that maximizes the likelihood. Lower inset: Assessed precision (variance $\delta^2$) of the three parameters (dots) compared to the $1/N^2$ scaling limits (solid lines) and the $1/N$ scaling bounds of the individual measurement strategy (dashed lines), extracted from $M=600$ sets of estimators, each derived from $n=600$ repeats of single-shot measurements. The definition of the error bars is described in Methods. The dashed arrow marks the reduction of $\delta \theta_{\rm{est}}$ over its classical bound. 
		}
	\end{center}
\end{figure*}

\bigskip

In the first experiment, we employ a maximally entangled state between the central module $\mathcal{A}$ and a single sensor module $\mathcal{B}$ to estimate a remote vector field. The protocol involves first creating a two-qubit entangled state locally on $\mathcal{A}$, then coherently transferring the state of one qubit in this pair to a sensor qubit on module $\mathcal{B}$. This distributed entangled state is subsequently used to perform simultaneous estimation of three independent components of a vector field at the sensor module. 

The second experiment uses two sensor modules, $\mathcal{B}$ and $\mathcal{C}$, to estimate multiple spatial gradients of vector fields. The protocol begins by locally preparing a two-qubit entangled state on the central module $\mathcal{A}$, followed by coherent transferring the state of each qubit to two different sensor modules. Local entangling gates are then applied on each sensor module to expand the state into a four-qubit distributed GHZ state. This entangled state is subsequently used to estimate multiple spatial gradients of vector fields applied at the sensor modules. Detailed descriptions of the experimental setup, calibration procedures, and implementation protocols are provided in Supplementary Information II.

For both protocols, we characterize the sensing performance of the distributed network across relevant parameter regimes, and estimate multiple parameters by applying maximum likelihood estimation (MLE) to measurement probabilities obtained from repeated single-shot measurements. 
This procedure enables a systematic evaluation of the sensing protocol’s overall performance, as well as the precision achievable in multi-parameter estimation.

\mysubsection{Sensing of remote vector fields}
 
\noindent
We first demonstrate the simultaneous estimation of three components of a remote vector field located at one sensor node. In this scenario, we realize a setup where the central module can perform measurements and entangled operations, whereas the sensor module, which interacts with the vector field, has limited capability and can only perform local operations.

The experiment procedure is illustrated in \rfig{fig2}{\bf a}. We begin by generating a Bell state $|\psi \rangle = \frac{1}{\sqrt{2}} (|00 \rangle + |11 \rangle)$ in the central module $\mathcal{A}$. One qubit, $Q_1$, is retained on $\mathcal{A}$, while the state of the second qubit, $Q_2$, is transferred to qubit $Q_3$ on the sensor module $\mathcal{B}$, where the vector field is located. This results in a distributed Bell state shared between $Q_1$ (on $\mathcal{A}$) and $Q_3$ (on $\mathcal{B}$), as detailed in Supplementary Information II.C. The sensor qubit $Q_3$ then interacts with the vector field via the signal unitary $U_s(\mathbf{x}) = e^{-i \mathbf{B} \cdot \boldsymbol{\sigma} T}$, where $\mathbf{x} = (B,\theta,\phi)$ are the spherical coordinates of $\mathbf{B}$ with $\mathbf{B} = (B\sin\theta \cos\phi, B\sin\theta \sin\phi, B\cos\theta)$ , $\boldsymbol{\sigma}$ is the vector of Pauli matrices, and $T$ is the signal encoding time. 
In our experiment, the signal unitary is implemented using a sequence of calibrated quantum gates to accurately emulate field-induced quantum evolution. This approach enables precise and controlled benchmarking of the sensing protocol under well-defined conditions. For practical applications such as microwave sensing, the vector field components may represent physical parameters including field amplitude, phase, and frequency detuning. 
Each application of the signal unitary is followed by a control unitary $U_c$ applied to the sensor qubit. This sequence is repeated $N$ times resulting in a total evolution described by $[U_c U_s(\mathbf{x})]^N \otimes I$. 
To enhance sensing precision, the control operation can be optimized. Theoretically, the optimal control is given by {\small$U_{c} = U_s^{\dagger}(\mathbf{x})$}~\cite{Hou2021, Yuan2016, Hou2019, Hou2020, Liu2023}. 
In practice, as the parameters are initially unknown, the control needs to be implemented adaptively as {\small$U_{c} = U_s^{\dagger}(\mathbf{x}_c)$}, where $\mathbf{x}_c=(B_c, \theta_c, \phi_c)$ are estimated values of field parameters that are iteratively refined using prior knowledge and accumulated measurement data. After the sensing sequence, the state of the sensor qubit $Q_3$ is transferred back to $Q_2$ in module $\mathcal{A}$ where a projective measurement on the Bell basis is performed. By repeating the experiment $n$ times, we obtain the  probabilities, $\{P_{00}$, $P_{01}$, $P_{10}$, $P_{11}\}$, of the measurement outcomes.  Maximum likelihood estimation is then applied to these probabilities to extract estimators of the field parameters. Repeating the entire procedure $M$ times yields $M$ independent estimators, enabling a statistical characterization of the estimation precision. This process can be similarly applied to estimate fields at other sensor modules, and it can be parallelized, facilitating scalable and efficient multi-parameter quantum sensing. 

\begin{figure*}[!htb]
	\begin{center}
		\includegraphics[width=0.9\textwidth]{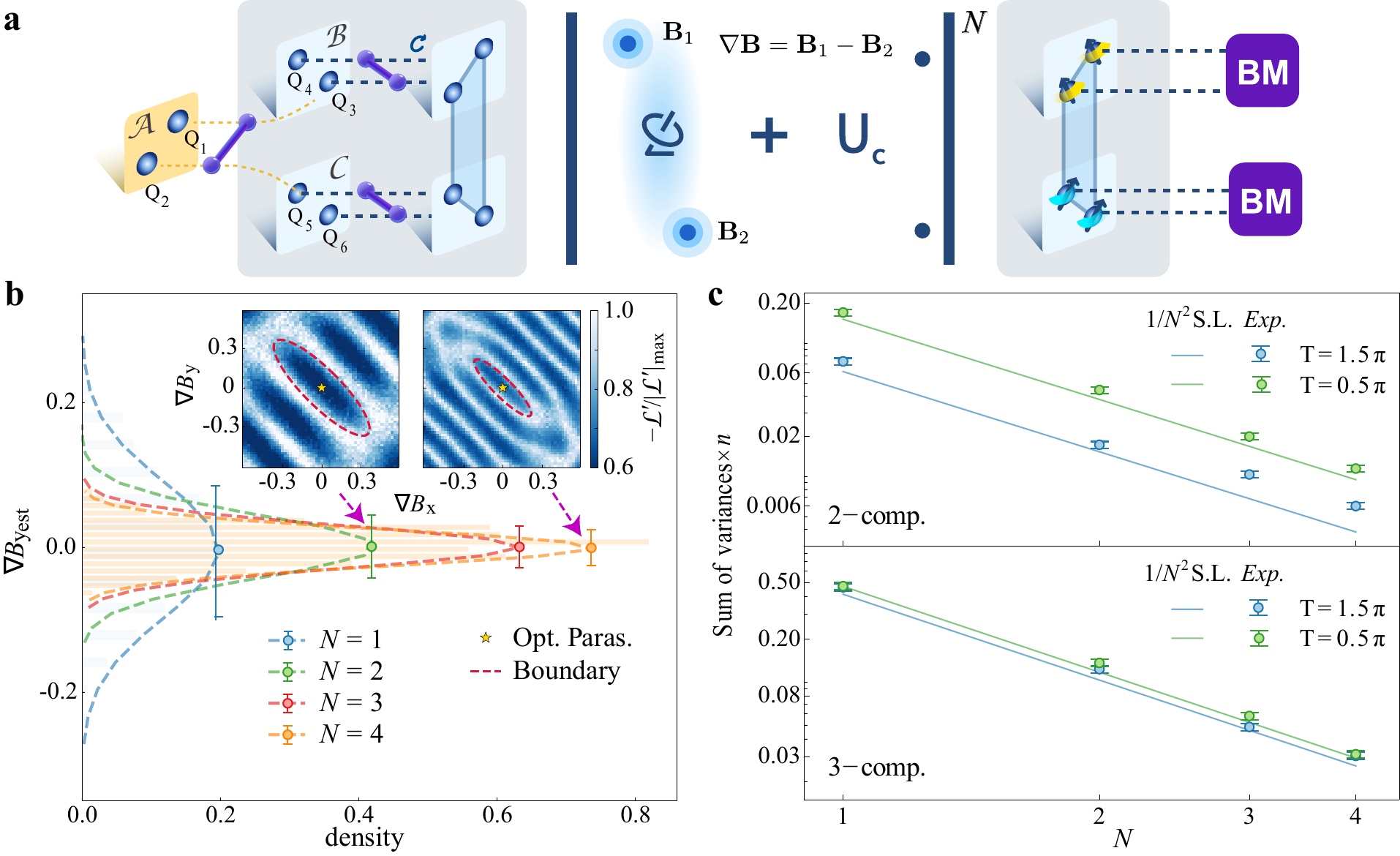}
		\caption{
			\label{fig3}
			{\bf Multi-parameter quantum metrology of the magnetic field gradient. }
               {\bf a,} Distributed sensing scheme for directly estimating the gradients of two vector fields with a non-local entangled state. Left section shows the probe state initialization. The system consists of a central module (yellow box) and two sensor modules (light blue boxes within the grey area). Blue spheres within the modules represent qubits, the purple stick-ball model represents CNOT gates, and the circular arrow denotes a local rotation. Four sensor qubits are initialized in a non-local entangled state, $ |\Psi_{0}\rangle = \frac{1}{\sqrt{2}}(|0011\rangle - |1100\rangle) $. The middle section illustrates the signal control encoding process, while the right section depicts simultaneous Bell measurements on the two encoded sensor modules. Sensor qubits in the same colored stripes are encoded with identical vector field signals.
			{\bf b,} Benchmarking the performance of the gradiometer.  Main panel: Distribution of estimator $\nabla B_{y_{\rm{est}}}$ for $N=1$ to $N=4$. Dashed lines represent the Gaussian fit of the estimator histograms, while circles and error bars indicate the mean values and standard deviations ($\delta \nabla B_{y_{\rm{est}}}$). Inset: Landscape of the log-likelihood function $\mathcal{L^{\prime}}$ at $N=2$ (left) and $N=4$ (right) (see Methods). The star marks the optimal control point and the red dashed contour denotes the parameter-estimation region. 
			{\bf c.} The gradient estimation precision, evaluated by the sum of variances, $\sum_{i \in \{x, y, z\} \cup \{x,y\}} \delta^2 \nabla B_{i_{\text{est}}}$, obtained from $M$ sets of estimators ($M=600$), each derived from $n$ measurement shots ($n=600$).
   Top panel: 2-component vector field. Bottom panel: 3-component vector field. Solid lines indicate the $1/N^2$ scaling limit ($1/N^2$ S.L.). Precision estimated at $T=0.5\pi$ is marked in green, while precision at $T=1.5\pi$ is marked in blue. The definition of the error bars is described in Methods.
   }
	\end{center}
\end{figure*}

\begin{figure*}[htb]
    \includegraphics[width=0.95\textwidth]{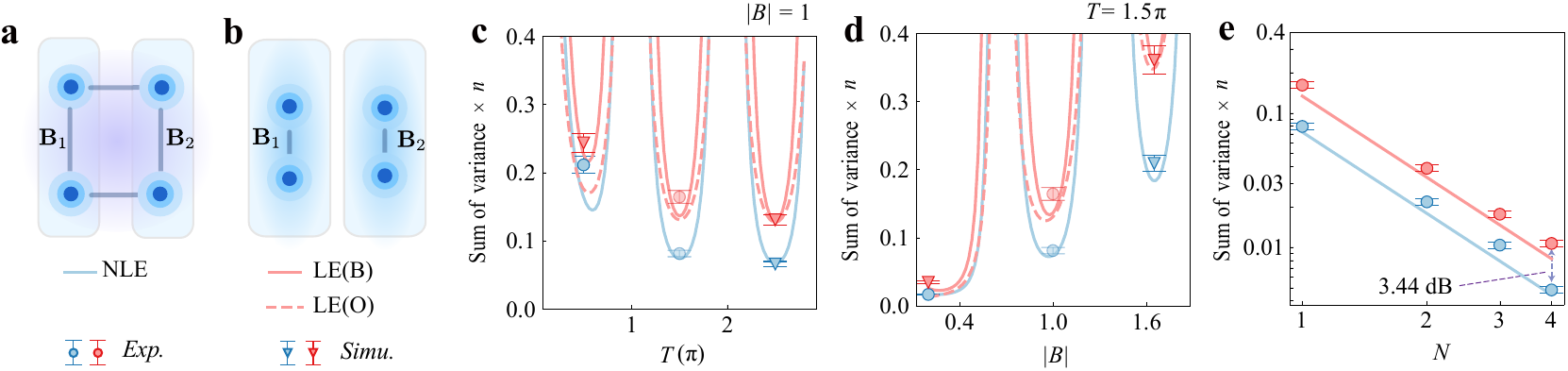}
    \caption{
        \label{fig4}
        {\bf Strategies comparison for gradient estimation of a 2-component vector field.}
    {\bf a-b,} Schematic diagrams of different strategies:
    ({\bf a}) Distributed sensing with non-local entanglement (NLE);
    ({\bf b}) Sensing with local entanglement (LE).  
    {\bf c-e,} Comparison of the precision ($\sum_{i\in \{x,y\}} \delta^2 \nabla\! B_{i_\text{est}}$) of the two strategies.
    ({\bf c}) Precision as a function of field strength $|B|$ at $T\ =\ 1.5\pi$ and $N=1$.
    ({\bf d}) Precision as a function of encoding time $T$ at $|B| = 1$ and $N = 1$.
    ({\bf e}) Precision versus number of cycles $N$ at $T = 1.5\pi$ and $|B| = 1$.
    The solid curve represents the theoretical precision bound for the local entanglement strategy with the probe state and  measurement taken as the Bell state and Bell measurement, labelled as LE(B). The dashed curve represents the theoretical precision bound for the local entanglement strategy with the optimal probe state and optimal measurement, labelled as LE(O). The definition of the error bars in (c)-(e) is described in Methods.
    }
\end{figure*}

In the experiment the signal parameters are set to $B=1$, $\theta=\frac{\pi}{4}$ and $\phi=\frac{\pi}{4}$. 
With optimal control in place, the total evolution simplifies to $U = I \otimes I$, meaning the final state remains identical to the initial Bell state. Consequently, the measured probability peaks at $P_{00}$. In \rfig{fig2}{\bf b}, we show the profile of $P_{00}$ as a function of the control parameters $B_c$ and $\phi_c$ for different values of the signal encoding time $T$. The width of the central peak around the true parameter values serves as a direct indicator of estimation precision. Notably, for a given circuit sequence depth $N$, the precision for each parameter does not indefinitely improve with increased $T$. The peak width for $B$ narrows, while for $\phi$ it oscillates as $T$ increases. At $T = 1.5\pi$, where the peak width for $\phi$ is the narrowest, we individually scan each control parameter. As shown in Fig. 2c, the central peak narrows consistently for all three parameters as $N$ increases, demonstrating a clear enhancement in estimation precision with greater sequence depth.

To assess measurement precision, we employ maximum likelihood estimation to obtain a set of estimated parameters $\mathbf{x}_{\text{est}} = (B_{\text{est}}, \theta_{\text{est}}, \phi_{\text{est}})$. For instance, \rfig{fig2}{\bf d} displays the distribution of the estimator $\phi_{\text{est}}$ for different values of $N$. As the number of cycles $N$ increases, the distribution narrows significantly, demonstrating the enhanced estimation precision achieved through the sequential control strategy. 
To further characterize the sensor performance throughout the parameter space, we experimentally construct two-dimensional likelihood landscapes by scanning the signal parameters $\theta$ and $\phi$ and recording the corresponding measurement probabilities (see \meth{sec:MLE} for details). These landscapes visually capture the structure of the likelihood function and illustrate how the estimator sharpens around the true parameter values with increasing $N$, as evidenced by the contraction of the high-likelihood region—consistent with a systematic improvement in estimation precision.

As shown in the lower inset of \rfig{fig2}{\bf d}, the estimation precision for all three parameters, $\{ B_{\text{est}}, \theta_{\text{est}}, \phi_{\text{est}} \}$, exhibits a $1/N^2$ scaling, matching the optimal precision achievable in corresponding single-parameter settings and indicating zero trade-off in simultaneous estimation of the three parameters. The enhancement is enabled by two key features: the use of a GHZ-type entangled probe state, which circumvents limitations imposed by locally non-commuting generators, and the adaptive control-enhanced sequential strategy, which amplifies precision through repeated signal-control interrogation cycles. For comparison, we benchmark these results against an individual measurement strategy that employs separable probes and measurements, where the total resources are divided evenly to estimate each parameter individually. The achieved maximum precision gains, in terms of variance, are 12.8 dB for $B_{\text{est}}$, 13.72 dB for $\theta_{\text{est}}$, and 12.56 dB for $\phi_{\text{est}}$.

\mysubsection{Distributed sensing of vector field gradients}
  
\noindent
In the second experiment, we realize a scenario where each sensor module is capable of performing local entangling operations and measurements. 
Our objective is to implement a distributed gradiometric protocol. Here, two sensor modules are exposed to distinct vector fields, $\mathbf{B}_1$ and $\mathbf{B}_2$, and we simultaneously estimate all components of the gradient $\nabla \mathbf{B} = \mathbf{B}_{1} - \mathbf{B}_{2}$.

We begin by preparing a non-local entangled (NLE) probe state $|\Psi_0\rangle = (|0011\rangle - |1100\rangle)/\sqrt{2}$ across sensor modules $\mathcal{B}$ and $\mathcal{C}$. The protocol starts with generating a Bell pair between qubits $Q_1$ and $Q_2$ in the central module $\mathcal{A}$, followed by coherent state transfer from $Q_1$ to $Q_3$ on $\mathcal{B}$ and from $Q_2$ to $Q_5$ on $\mathcal{C} $. Local CNOT gates are then applied within each sensor modules to extend the entanglement, resulting in a four-qubit GHZ state. The final probe state $|\Psi_0\rangle$ is obtained by applying additional X gates to $Q_3$ and $Q_4$, and a Z gate to $Q_5$. The overall preparation fidelity is measured to be $76.16\%$.

The entangled probe state then interacts with local vector fields $\mathbf{B}_1$ and $\mathbf{B}_2$. The evolution over time $T$ is governed by the operator $U_{\text{S}}(\mathbf{B}_1,\mathbf{B}_2) = U_{s1}^{\otimes 2} \otimes U_{s2}^{\otimes 2}$, where $U_{sj} = e^{-i \mathbf{B}_{j} \cdot \boldsymbol{\sigma} T}$ for $j = 1, 2$. This evolution can be reparametrized as $U_{\text{S}}(\nabla \mathbf{B}, \sum \mathbf{B})$, where $\nabla \mathbf{B} = \mathbf{B}_1 - \mathbf{B}_2$ represents the gradient and $\sum \mathbf{B} = \mathbf{B}_1 + \mathbf{B}_2$ represents the sum. 
After each signal evolution, a control operation $U_{\text{C}}$ is applied. The cycle is repeated $N$ times, resulting in the total evolution $[U_{\text{C}}U_{\text{S}}(\nabla \mathbf{B}, \sum \mathbf{B})]^N$. The control operation is implemented adaptively as {\small$U_{\rm{C}} = U_\rm{S}^{\dagger}(\nabla \mathbf{B}_C,\sum \mathbf{B}_C)$} where $(\nabla \mathbf{B}_C, \sum \mathbf{B}_C)$ are the iteratively updated estimates of $(\nabla \mathbf{B}, \sum \mathbf{B})$ based on accumulated measurement outcomes. 
Finally, Bell measurements are performed on both modules $\mathcal{B}$ and $\mathcal{C}$. 
The experiment is repeated $n$ times to obtain the probability distribution $\{P_{ijkl}\}$, where $i,j,k,l \in \{0,1\}$. From these probabilities, the gradients $\nabla \mathbf{B}_{\text{est}} = (\nabla B_x, \nabla B_y, \nabla B_z)$ are inferred via maximum likelihood estimation.

To validate this protocol, we examine two representative configurations for the vector fields $\mathbf{B}_1$ and $\mathbf{B}_2$. The first is $\mathbf{B}_1 = \mathbf{B}_2 = \sqrt{2}/4(1,1,0)$, corresponding to two vector fields confined to the XY-plane with a known zero Z-component. The second is a fully three-dimensional case with $\mathbf{B}_1 = \mathbf{B}_2 = 1/2(1,1,\sqrt{2})$. We assume $\mathbf{B}_1 = \mathbf{B}_2$ in both cases without loss of generality, since any initial mismatch $\mathbf{B}_1 \neq \mathbf{B}_2$ can be iteratively corrected by applying an adaptive compensation $\nabla \mathbf{B}_{\text{est}}$ to one of the nodes. This procedure asymptotically aligns the two fields and does not affect the ultimate estimation performance~\cite{pang2017optimal} (see Supplementary Information I.C.1). 
For both configurations, experiments are conducted for $N = 1$ to 4 cycles. As shown in \rfig{fig3}{\bf c}, the observed precision approaches the theoretical limit and exhibits the expected $1/N^2$ scaling. Further increases in $N$ are currently constrained by decoherence and control errors. We quantitatively analyze the impact of these noise sources and their contribution to deviations from the theoretical limit in Supplementary Information III.C, and note that such limitations can be mitigated through future hardware improvements.

Under optimal control unitary, the probabilities of the measurement results are dominated by $P_{0010}$ and $P_{1000}$ (see Supplementary Information II.C). 
Taking the $y$-component estimator $\nabla B_{y_{\rm{est}}}$ for the 2D gradient case as an example, its normalized distribution in \rfig{fig3}{\bf b} shows that as $N$ increases from 1 to 4 at $T=1.5\pi$, the standard deviation progressively decreases, indicating improved estimation precision.  Similarly, the precision for the $x$-component, $\delta \nabla B_{x_{\rm{est}}}$, follows an identical trend, see Supplementary Information II.C and Supplementary Information III.B. 

For comparison, we also perform an experiment using only local entanglement within each module (see \rfig{fig4}{\bf a,b}). In this reference scheme, both qubits in each sensor module serve as local sensors. Bell states are prepared independently within modules $\mathcal{B}$ and $\mathcal{C}$ to estimate $\mathbf{B}_1$ and $\mathbf{B}_2$ separately. In each module, the signal encoding is interleaved with optimal control operations, and Bell measurements are performed after the evolution. The gradient $\nabla \mathbf{B}$ is then computed by differencing the two local estimates. We use {\small $g_\text{LE/NLE} = 10\log_{10}[(\sum_{i\in \{x,y\}} \delta^2 \nabla\! B_{i_\text{est}})_\text{LE}^\text{exp(sim)}/(\sum_{i\in \{x,y\}} \delta^2 \nabla\! B_{i_\text{est}})_\text{NLE}^\text{exp} ]$} to quantify the advantage provided by distributed sensing with non‑local entanglement (NLE) over the local sensing with local entanglement (LE). Similar comparisons between local and non-local strategies have also been made in prior works~\cite{PhysRevLett.121.130503,Liu2020}. Here, we restrict the comparison to the two‑component gradient, because in the three‑component case the quantum Fisher information matrix for the local strategy becomes singular, rendering the estimation unfeasible for the local sensing (see Supplementary Information I.C.2). We assess the gain in the two-component case under several conditions: first, by fixing the field amplitude at  $|\mathbf{B}| = 1$ and sweeping the encoding time $T$ from $0.5\pi$ to $2.5\pi$ (\rfig{fig4}{\bf c}); next, by fixing $T = 1.5\pi$ and scanning $|\mathbf{B}|$ over the range 0.2 to 1.65 (\rfig{fig4}{\bf d}); and finally, by holding $|\mathbf{B}| = 1$ and $T = 1.5\pi$ constant while varying the number of sequential encoding steps $N$. For $N = 4$, the distributed protocol achieves a maximal gain of 3.44 dB over the local strategy.

\mysection{Discussion}

\noindent
By deterministically generating high-fidelity distributed entanglement across network nodes, our experiment demonstrates an an inverse-square scaling of the variance with sensing circuit depth for the estimation of all three components of a remote vector field, with an improvement of up to 13.72 dB over individual strategy. In the context of gradient estimation, our distributed strategy---enabled by modular superconducting architecture, distributed entanglement, and control-enhanced sequential strategies---surpasses local strategies, achieving a 3.44 dB reduction in total variance when estimating gradients along two distinct directions in distributed two-dimensional vector fields. 
The four-qubit GHZ state between two non-nearest-neighbor nodes used in the gradient sensing exhibits a fidelity of $80.36\%$, which is among the highest demonstrated across physically separated network nodes to date~\cite{Niu2023,ruskuc2025multiplexed, pompili2021realization, main2025distributed} and sufficient to realize a clear advantage of the distributed protocol.

These results establish a scalable framework for distributed quantum sensing of vector fields and their gradients with enhanced precision and architectural flexibility. Our approach offers a concrete path toward quantum-enhanced sensor networks that can be applied to a wide range of practical scenarios, including electromagnetic field monitoring, navigation, and remote detection. Looking forward, the integration of adaptive control, error correction, and expanded network topologies could unlock new frontiers in precision sensing and real-time quantum signal processing.


\mysection{Methods}

\begin{figure*}[htb]
    \includegraphics[width=0.75\textwidth]{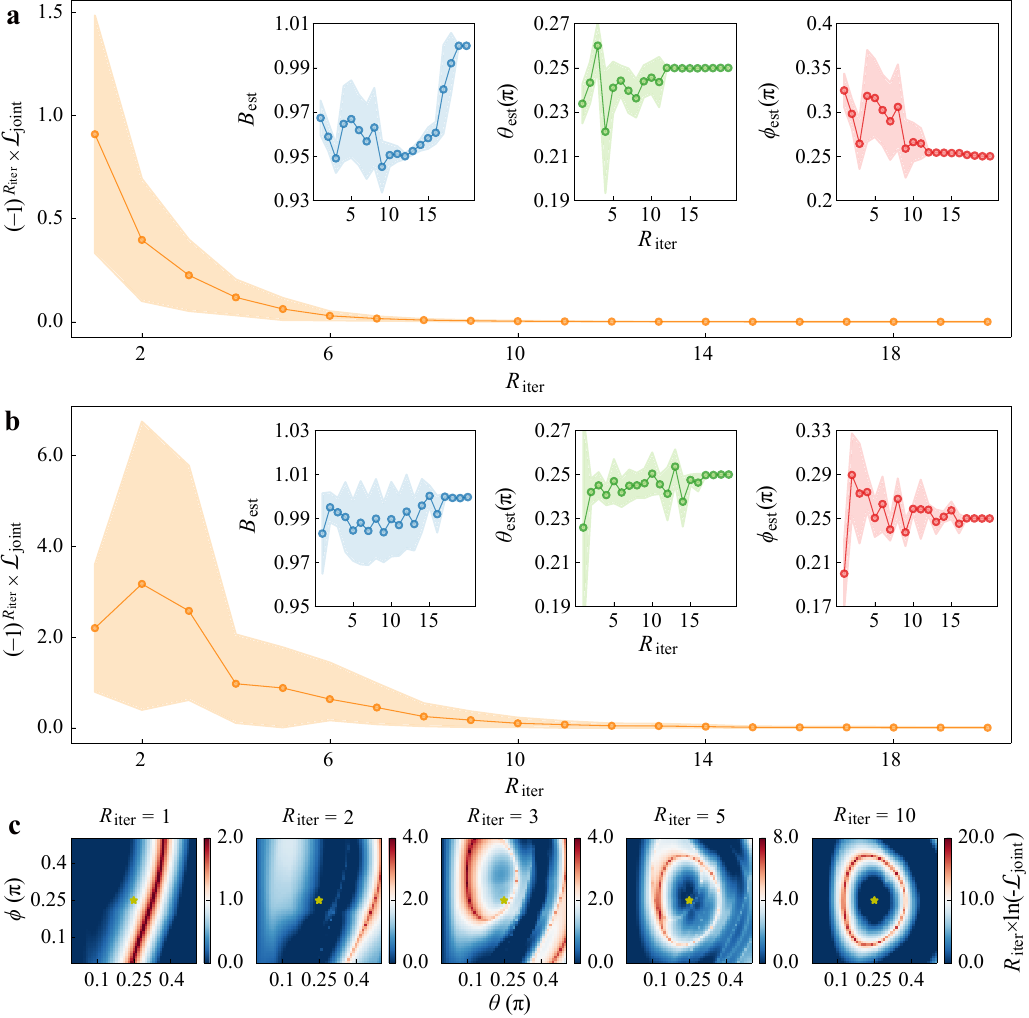}
    \caption{
        \label{fig5}
        {\bf Adaptive control-enhanced metrology for simultaneous three-parameter estimation with signal parameters $(B,\theta,\phi) = (1, \frac{\pi}{4},\frac{\pi}{4})$.}
        Here $R_\text{iter}$ denotes the iteration index of the adaptive estimation-control loop. 
        {\bf a,} The results of adaptive iterations starting from different initial guess values (10 sets randomly chosen within the boundary of the landscape) with $N=1$. 
        {\bf b,} The results of adaptive iterations with different sequential copies $N=1$ to $6$ and fixed initial guess (1 set randomly chosen within the boundary of the landscape for $N=1$). 
        {\bf c,} The calculated likelihood function landscape for $N=4$ after iteration cycle $1,2,3,5,10$. Stars indicate the locations of the optimal control parameters.
         }
\end{figure*}

\mysubsection{Experimental platform}\label{sec:setup}

\noindent
We implement the distributed quantum sensor on a modular superconducting quantum network consisting of five modules~\cite{Niu2023}. Each module hosts four capacitively coupled transmon qubits, enabling local entangled operations. Inter-module communication is realized via high-fidelity quantum state transfer through low-loss microwave links (four 25-cm aluminum coaxial cables). Tunable couplers at each cable–qubit interface allow programmable interaction between the qubits and the multimode cable resonators, supporting coherent photon transfer across modules. The entangled probe states are prepared through a combination of quantum state transfer, local CNOT gates, and single-qubit rotations (see Supplementary Information II.C). The vector field signal and control unitaries are digitally simulated using a $U(3)$ formalism (see Supplementary Information II.B), where the signal parameters are set to span representative conditions, while control parameters are set according to prior experimental calibration. To extract information about the encoded signals parameters, Bell measurements are performed on selected qubits to obtain the output state probabilities. Maximum likelihood estimation is then used to reconstruct the parameters of interest from the measurement data, enabling quantitative evaluation of sensor precision and performance.

\mysubsection{Maximum likelihood estimation for remote field sensing}\label{sec:MLE}

\noindent
To estimate the unknown signal parameters, we employ maximum likelihood estimation (MLE) based on experimentally acquired measurement outcomes~\cite{casella2024statistical}. For a fixed control configuration $\mathbf{x}_c$, we implement the sensing circuit and perform $n$ repeated measurements on a fixed basis (e.g., the Bell basis), yielding outcomes $y_i \in \{00,01,10,11\} (j=1,\cdots, n)$. Let $n_i$ be the count of outcome $i \in \{00,01,10,11\}$ and $P_i^{\rm{exp}} = n_i/n$ the corresponding empirical frequency. Denote by $P_{i}^{\rm{ideal}}(\mathbf{x},\mathbf{x}_c)$ the ideal model-predicted probability of outcome $i$ under parameter $\mathbf{x}$ and control $\mathbf{x}_c$. 
For clarity, we first define the likelihood of the full measurement record
\begin{equation}
\mathcal{L}(y_1,\ldots,y_n\,|\,\mathbf{x},\mathbf{x}_c)=\prod_{j=1}^{n} P^{\rm{ideal}}(y_j\,|\,\mathbf{x},\mathbf{x}_c).
\end{equation}
Grouping identical outcomes, we define the normalized log-likelihood objective as
\begin{equation}
    \begin{aligned}
        \mathcal{L}(\mathbf{x},\mathbf{x}_{\rm{c}}) \equiv& \frac{1}{n}\ln \mathcal{L}(y_1,\ldots,y_n\,|\,\mathbf{x},\mathbf{x}_c)\\
=& \sum_i P_{i}^{\rm{exp}} \ln\!\big(P_{i}^{\rm{ideal}}(\mathbf{x},\mathbf{x}_c)\big).
    \end{aligned}
\end{equation}
The estimated parameters are obtained by maximizing the normalized log-likelihood function: $\mathbf{x}_{\text{est}} = \mathop{\arg\max}\limits_{\mathbf{x}} \mathcal{L}(\mathbf{x},\mathbf{x}_{\rm{c}})$, using gradient-based optimization. To quantify the estimation precision, we perform $M$ independent realizations of the full estimation procedure to obtain $M$ independent estimators for each parameter, from which the variance $\delta^2$ is evaluated; the error bars of the plotted variances in Figs.\ref{fig2},\ref{fig3},\ref{fig4} are then given by its standard deviation $\mathrm{SD}(\delta^2) = \sqrt{2/(M-1)}\delta^2$~\cite{ahn2003standard}.

Since the optimal control parameters depend on the true signal parameters, we implement an adaptive protocol to iteratively refine $\mathbf{x}_c$~\cite{Yang2021,Kaubruegger2019,Marciniak2022,pang2017optimal}. This process consists of four steps.
1) Initialization: A set of control parameters, \small{$\mathbf{x}_c^{(0)} = ( B_c^{(0)}, \theta_c^{(0)}, \phi_c^{(0)})$}, is chosen (e.g., randomly or based on prior knowledge). The sensing circuit is executed, yielding the empirical probability distribution \small{$\{P_i^{\rm{exp(0)}}\}$}. 
2) First estimation: The first parameter estimate is obtained via MLE: \small{ $\mathbf{x}_{\text{est}}^{(1)} = \mathop{\arg\max}\limits_{\mathbf{x}} \sum_i P_i^{\rm{exp(0)}} \ln  (P_i^{\rm{ideal}}(\mathbf{x}, \mathbf{x}_c^{(0)}) )$}. This estimate is then used to update the control parameters \small{$\mathbf{x}_c^{(1)} = \mathbf{x}_{\text{est}}^{(1)}$}. 
3) Iteration: Using the updated control parameters $\mathbf{x}_c^{(1)}$, the experiment is repeated to obtain a new empirical distribution ${P_i^{\rm{exp(1)}}}$. 
4) Joint Estimation: The estimator and control parameters are updated by maximizing the joint log-likelihood incorporating all data collected so far:
\small{$\mathbf{x}_{c}^{(2)} =\mathbf{x}_{\text{est}}^{(2)} = \mathop{\arg\max}\limits_{\mathbf{x}} \sum_{i} P_i^{\rm{exp(0)}} \ln ( P_i^{\rm{ideal}}(\mathbf{x}, \mathbf{x}_c^{(0)}) ) \times \sum_{i} P_i^{\rm{exp(1)}} \ln ( P_i^{\rm{ideal}}(\mathbf{x}, \mathbf{x}_c^{(1)}) )$}. 
This adaptive process is repeated for $K$ cycles. The final estimate after $K$ iterations is:
\small{$\mathbf{x}_{\text{est}}^{(K)} = \mathop{\arg\max}\limits_{\mathbf{x}} \mathcal{L}_{\text{joint}}$}, where \small{$\mathcal{L}_{\text{joint}} = \prod_{m=1}^{K-1}\sum_{i} P_{i}^{\rm{exp}(m)} \ln[P_{i}^{\rm{ideal}}(\mathbf{x},\mathbf{x}_c^{(m)})]$} is the joint likelihood function~\cite{Clausen2024}.
The $K$-th MLE considers all $K-1$ experiment results, the control parameters will be updated with the increasing of iteration cycles. After at most $R_\rm{iter} = 40$ rounds,
the estimators $B_{\rm{est}}$, $\theta_{\rm{est}}$, and $\phi_{\rm{est}}$ converge to stable values.

We demonstrate the convergence of the adaptive protocol in \rfig{fig5} by minimizing the joint likelihood function $-\mathcal{L}_{\text{joint}}$ in each cycle and tracking the evolution of the cost function $(-1)^{R_{\rm{iter}}} \times \mathcal{L}_{\text{joint}}$. The convergence process is shown to be robust both across various initial guesses at $N=1$ and across sequential copies ($N=1$ to $6$) for a fixed initial guess, as depicted in \rfig{fig5}{\bf a},{\bf b}. Numerical analysis of the joint likelihood function at $N=4$ (\rfig{fig5}{\bf c}) indicates that, while the landscape initially appears irregular, it evolves into an optimal configuration similar to that in \rfig{fig2}{\bf b} after a few cycles, highlighting the effectiveness of this adaptive optimization process.

\mysubsection{Experimental likelihood benchmark}

\noindent
To assess the sensitivity and precision of our sensing protocol under a given control setting, we perform a likelihood function benchmark. It is crucial to distinguish this from the parameter estimation procedure: this benchmark is not used to extract unknown parameters. Instead, its purpose is to visualize the structure of the likelihood landscape and validate the agreement between our experimental results and the theoretical model.

This benchmarking method operates as follows: We fix the control parameters $\mathbf{x}_c$ to their optimal values and then scan the signal parameters $\mathbf{x}$ over a local grid. For each point on this grid, we run the sensing experiment and record the resulting measurement probability distribution ${ P_{i}^{\rm{exp}}(\mathbf{x},\mathbf{x}_c) }$. We then construct the experimental log-likelihood function $\mathcal{L}^\prime = \sum_i P_{i}^{\rm{exp}} \ln(P_{i}^{\rm{exp}}(\mathbf{x},\mathbf{x}_c))$ by comparing these probabilities to the reference distribution ${ P_{i}^{\rm{exp}} }$ measured at the true signal parameter values.

The 2D landscapes of the normalized $\mathcal{L}^\prime$, shown in Fig. 2d and Fig. 3b, reveal the parameter estimation space. This space is characterized by an internal boundary whose saddle point corresponds to the alignment of the guess signal parameters with the true values. The contraction of this boundary with an increasing number of cycles $N$ visually demonstrates the enhanced precision achieved through our sequential strategy.

In summary, this benchmark provides an intuitive, fully data-driven method to evaluate sensing performance; it is strictly a diagnostic tool and does not influence the adaptive protocol or the final estimation outcome. The close agreement between our experimentally probed landscapes and numerical simulations validates the underlying model used in our actual MLE procedure (see Supplementary Information III.A). This paves the way for developing a programmable sensor system reliant solely on direct experimental measurements and parameter feedback, with the benchmark serving as a crucial verification step.

\mysubsection{Ultimate precision for remote vector field sensing}

\noindent
The Hamiltonian for the sensor qubit can be expressed as {\small$H = \mathbf{B}\cdot \boldsymbol{\sigma}$}, where the vector field $\mathbf{B} = (B_x, B_y, B_z)$ can be expressed in the spherical coordinates as $B_x = B\sin\theta \cos\phi$, $B_y = B\sin\theta \sin\phi$, and $B_z = B\cos\theta$ with $\mathbf{x} = (B, \theta, \phi)$ as the parameters in the spherical coordinates. The unitary operator generated under the free evolution over time $T$ is given by $U_s(\mathbf{x}) = e^{-i \mathbf{B} \cdot \boldsymbol{\sigma} T}$. 

To evaluate the performance of a sensing strategy, we employ the quantum Cram\'er-Rao bound, given by $n\text{Cov}(\mathbf{x}_{\text{est}}) \geq F_Q^{-1}$, where $\text{Cov}(\mathbf{x}_{\text{est}})$ is the covariance matrix of the estimators for the unknown parameters $\mathbf{x}$, $n$ is the number of measurement repetitions, and $F_Q$ is the quantum Fisher information matrix (QFIM). The overall estimation precision for multiple parameters is quantified by the total variance $(\operatorname{Tr}[\operatorname{Cov}(\mathbf{x}_{\text{est}})])$.

For the optimally controlled strategy, the optimal control $U_c = U_s^\dagger$ is applied after each signal unitary $U_s$, and the sequence is repeated $N$ times~\cite{Hou2021, Yuan2016, Hou2019, Hou2020, Liu2023}. In this case, the quantum Fisher information matrix is given by {\small$F_Q^{\max} = 4 N^2 \left(\begin{array}{ccc}
T^2 & 0 & 0 \\
0 & \sin^2 (B T) & 0 \\
0 & 0 & \sin^2 (B T) \sin^2 \theta
\end{array}\right)$}. 
In Supplementary Information I.B, we also explicitly compute the classical Fisher information matrix (CFIM) under the projective measurements in the Bell basis, which matches the QFIM, confirming that the quantum Cramér-Rao bound can be saturated. The ultimate precision limits for the three parameters are then given by 
$n (\delta B_{\text{est}})^2 \ge \frac{1}{4 N^2 T^2}$, 
$n (\delta \theta_{\text{est}})^2 \ge \frac{1}{4 N^2 \sin^2(BT)}$, 
$n (\delta \phi_{\text{est}})^2 \ge \frac{1}{4 N^2 \sin^2(BT)\sin^2\theta}$, where $(\delta x_{\text{est}})^2$ denotes the variance of the estimator and $n$ is the number of measurement repetitions. 
For each of the parameters, this is also the highest precision that can be achieved in the single-parameter case where the other two parameters are taken as known values. The optimal strategy thus achieves the highest precision for all three parameters simultaneously without any tradeoff.

This is compared with the individual measurement strategy \cite{Hou2021}, where the total number of channel uses $N$ is divided evenly among three groups, and each group is dedicated to estimating one parameter independently. This strategy exhibits a $1/N$ scaling with respect to the number of channel uses as 
$n (\delta B_{\text{est}})^2 \ge \frac{3}{4 N T^2}$, 
$n (\delta \theta_{\text{est}})^2 \ge \frac{3}{4 N \sin^2(BT)}$, 
$n (\delta \phi_{\text{est}})^2 \ge \frac{3}{4 N \sin^2(BT)\sin^2\theta}$. 
This serves as a baseline for assessing the enhancements provided by the distributed strategy.

\mysubsection{Precision limit for gradient estimation with non-local entanglement}

\noindent
Under the optimal controlled sequential scheme, the total dynamics is given by {\small$(U_{\rm{C}} U_{\rm{S}})^N $}, where {\small$U_{\rm{S}}(\mathbf{x}) = U_{s1}^{\otimes 2} \otimes U_{s2}^{\otimes 2}$}, {\small$U_{\rm{C}} = U_{\rm{S}}^{\dagger}(\mathbf{x}_c)$} with {\small $\mathbf{x}_c = (\nabla \mathbf{B}_{c}, \sum\!\mathbf{B}_{c})$} as the adaptively updated estimate of $\mathbf{x}=(\nabla \mathbf{B}, \sum\!\mathbf{B})$~\cite{Hou2021, Yuan2016}.
As detailed in the Supplementary Information I.C.1, using a non-local entangled probe state $|\Psi_{0} \rangle= (|0011\rangle - |1100\rangle)/\sqrt{2}$, the QFIM for the simultaneous estimation of $\mathbf{x}$ is given by
{\small$F_Q = N^2 \left(\begin{array}{cc}
F_{-} & \boldsymbol{0}  \\
\boldsymbol{0} & F_{+}
\end{array}\right)$}, 
where $F_{-}$ and $F_{+}$ correspond to the QFIMs for estimating {\small $\nabla \mathbf{B}$} and {\small$\sum\!\mathbf{B}$}, respectively. 
We further verify that under local projective measurements in the Bell basis on each sensor module, the classical Fisher information matrix coincides with the QFIM, confirming that the quantum Cramér-Rao bound can be saturated by the measurement strategy.
The block diagonal form of the QFIM implies that the precision of estimating the gradients {\small$\nabla \mathbf{B}$} is not affected by {\small$\sum\!\mathbf{B}$}. 
For the estimation of  {\small$\nabla B_x$} and {\small$\nabla B_y$} of two-dimensional vector fields, we have
\begin{seqnarray}
\begin{split}
\label{eq:pre_non-local}
    &n \left(\left(\delta \nabla B_{x_{\text {est }}}\right)^2+\left(\delta \nabla B_{y_\text{est }}\right)^2 \right)\\
    \geq &\frac{1}{4 N^2}\left( \frac{1}{T^2}+\frac{B^2}{\left(1+3 \sin ^2(B T)\right) \sin ^2(B T)}\right),
\end{split}
\end{seqnarray}\noindent
where $B$ is the magnitude of the vector field with {\small$B = \sqrt{B_x^2 + B_y^2 + B_z^2}$}. For the analysis of the three-dimensional case, see Supplementary Information I.C.1.

\mysubsection{Precision limit for gradient estimation with local entanglement}

\noindent
We theoretically determine the precision limits for the estimation of the gradients under the local strategy. In this strategy, local entangled states are employed to first estimate the vector field at each sensor module separately; the gradients are then obtained by computing the differences. Under this strategy, the precision for estimating {\small$\nabla B_x$} and {\small$\nabla B_y$} of two-dimensional vector fields is given by
\begin{sequation}
    \label{eq:pre_local_optimal}
        n \left(\left(\delta \nabla B_{x_{\text {est }}}\right)^2+\left(\delta \nabla B_{y_{\text {est }}}\right)^2 \right)\geq  \frac{1}{8 N^2}\left(\frac{1}{T^2}+\frac{B^2}{\sin ^2(B T)}\right).
\end{sequation}\noindent
This, however, requires an optimal probe state that depends on the true values of the parameters, which is unknown a priori. An adaptive preparation is thus needed. For practical implementation, we use the Bell state $\frac{1}{\sqrt{2}}(|00\rangle + |11\rangle)$ instead, which is parameter-independent and does not require adaptive preparation. In this case, the precision is given by 
\begin{sequation}
    \begin{aligned}
    \label{eq:pre_local_bell}
        &n\left(\left(\delta \nabla B_{x_{\text {est }}}\right)^2+\left(\delta \nabla B_{y_{\text {est }}}\right)^2 \right)\\ \geq& \frac{1}{8 N^2 \sin ^2(B T)}\left(\frac{B^2-\cos ^2(B T) B_x^2}{T^2 B_x^2}+\frac{B^2}{\sin ^2(B T)}\right).
    \end{aligned}
\end{sequation}\noindent
When $B_x = B_y$, as the case in the experiment, the two bounds become almost the same near the optimal time point with $\sin(B T) = 1$.

In both cases, the corresponding precision bound can be saturated by performing projective measurements in the Bell basis on each sensor module. We explicitly calculate the CFIM under the projective measurement of Bell-basis, confirming the optimality of the measurement strategy in our system.
The full derivation is provided in Supplementary Information I.C.2.

\clearpage


\mysection{Acknowledgements}

\noindent
We thank Xiu-Hao Deng, Zhibo Hou and Raphael Kaubruegger for insightful discussions. 
This work was supported by the National Natural Science Foundation of China (12374474 and 12174178), the Quantum Science and Technology-National Science and Technology Major Project (2021ZD0301703), the Science, Technology and Innovation Commission of Shenzhen Municipality (KQTD20210811090049034, RCBS20231211090824040, RCBS20231211090815032), the Shenzhen-Hong Kong Cooperation Zone for Technology and Innovation (HZQB-KCZYB-2020050), Guangdong Basic and Applied Basic Research Foundation (2024A1515011714, 2022A1515110615), Research Grants Council of Hong Kong (14309223, 14309624, 14309022), the Guangdong Provincial Quantum Science Strategic Initiative (GDZX2303007), and the Department of Science and Technology of Guangdong Province (2020B0303050001).

\mysection{Author contributions}

\noindent
J.N. initiated the project and designed the experiment.
J.J.Z. conducted the measurements and analyzed the data with Y.-J.H. under the supervision of J.N. 
L.W. and Y.-J.H. provided theoretical support guided by H.Y. 
J.W.Z. developed the microwave electronics infrastructure  with assistance from X.S. 
L.Z. fabricated the devices with support from Y.X.Z and S.L. 
W.H., J.Q. and Y.L. contributed to the experimental setup. J.C., J.J., Z.T., Y.C., X.L., W.G. and W.R. participated in discussions of the results.
J.J.Z., L.W., Y.-J.H., J.N. and H.Y. wrote the manuscript with input from all authors.
H.Y., J.N., Y.P.Z. and D.Y. supervised the project.

\mysection{Competing interests}

\noindent
The authors declare no competing interests.

\mysection{Data availability}

\noindent
The data that support the plots within this paper and other findings of this study are available from the corresponding author upon request.

\clearpage

\end{document}


\title{Supplementary Information for ``Distributed multi-parameter quantum metrology with a superconducting quantum network''}

\maketitle

\setcounter{equation}{0}
\setcounter{figure}{0}
\setcounter{table}{0}

\renewcommand{\theequation}{S\arabic{equation}}
\renewcommand{\thefigure}{S\arabic{figure}}
\renewcommand{\thetable}{S\arabic{table}}
\newcommand*{\bluenp}{\textcolor{bluenp}}
\renewcommand{\figurename}{Fig.}

\renewcommand{\figurename}{Supplementary Figure}
\renewcommand{\tablename}{Supplementary Table}

\newcommand{\rSupeq}[1]{Supplementary Equation~\ref{#1}}
\newcommand{\rSupTab}[1]{Supplementary Table~\ref{#1}}
\newcommand{\rSupTabs}[1]{Supplementary Tables~\ref{#1}}
\newcommand{\rSupFig}[1]{Supplementary Figure~\ref{#1}}
\newcommand{\rSupFigs}[1]{Supplementary Figures~\ref{#1}}

\tableofcontents
\newpage


\section{Theoretical analysis}

\subsection{General process of quantum metrology}

The primary objective of quantum metrology is to precisely estimate unknown physical quantities by utilizing quantum resources, such as quantum entanglement. The general process of quantum metrology includes the following steps: state preparation, parameter encoding, measurement, and estimation. The probe state $\rho_0$ evolves under the given dynamics, which depend on the unknown parameters $\mathbf{x}=(x_1, \cdots, x_n)$, resulting in the encoded state $\rho(\mathbf{x})$. To extract information about the parameters $\mathbf{x}$, we perform a set of positive operator-valued measures (POVMs), represented as $\{\Pi_i\}$, on the state $\rho(\mathbf{x})$, obtaining a set of probability distributions ${P_i(\mathbf{x})}$, where $P_i(\mathbf{x})$ is the probability of obtaining the measurement result $i$. Finally, we construct the estimators $\mathbf{x}_{\text{est}}=(x_{1_{\text{est}}}, \cdots, x_{n_{\text{est}}})$ based on the probabilities of the measurement outcomes. 
For multi-parameter quantum estimating, the performance of locally unbiased estimators is quantified by the covariance matrix, where the $jk$-th element gives
\begin{sequation}
	[\text{Cov}(\mathbf{x}_{\text {est}})]_{jk} = E\left[(x_{j_{\text{est}}}-x_j)(x_{k_{\text{est}}}-x_k)\right].
\end{sequation}\noindent
The estimation precision for multiple parameters is quantified by the sum of variances, which corresponds to the sum of the diagonal terms of the covariance matrix. The covariance matrix is lower bounded by
\begin{sequation}
	\text{Cov}(\mathbf{x}_{\text {est}}) \geq \frac{1}{n} F_C^{-1}
\end{sequation}\noindent
which is known as the Cramér-Rao bound. Here $n$ is the number of measurement repetitions, and $F_C$ is the Fisher information matrix (FIM) with $jk$-th element calculated as follows:
\begin{sequation}
	[F_C]_{x_j x_k} = \sum_i \frac{1}{P_i(\mathbf{x})} \left(\frac{\partial P_i(\mathbf{x})}{\partial x_j}\right)  \left(\frac{\partial P_i(\mathbf{x})}{\partial x_k}\right)
 \label{eq:cfim}
\end{sequation}\noindent
The Cramér-Rao bound is achievable for a large number of repetitions by using the maximum likelihood estimator (MLE). 
The quantum Cramér-Rao bound (QCRB) further constrains the covariance matrix:
\begin{sequation}
	\text{Cov}(\mathbf{x}_{\text {est}}) \geq \frac{1}{n} F_C^{-1} \geq \frac{1}{n} F_Q^{-1}.
\end{sequation}\noindent
Here $F_Q$ is the quantum Fisher information matrix (QFIM). For parameters encoded in pure states $|\psi_{\mathbf{x}}\rangle$, $F_Q$ can be expressed as:
\begin{sequation}
	[F_Q]_{x_j x_k} = 4 \operatorname{Re} \left(\left\langle\partial_{x_j} \psi_{\mathbf{x}} | \partial_{x_k} \psi_{\mathbf{x}}\right\rangle-\left\langle\partial_{x_j} \psi_{\mathbf{x}} | \psi_{\mathbf{x}}\right\rangle\left\langle\psi_{\mathbf{x}} | \partial_{x_k} \psi_{\mathbf{x}}\right\rangle\right).
	\label{eq:qfim_pure}
\end{sequation}\noindent
More specifically, consider a pure probe state $|\psi_0\rangle$ undergoes a unitary process $U_{\mathbf{x}}$, the encoded state $|\psi_{\mathbf{x}}\rangle = U_{\mathbf{x}} |\psi_0\rangle$. Define the generator of $U_{\mathbf{x}}$ corresponding to unknown parameter $x_j$ as 
\begin{sequation}
	h_{x_j} = i U_{\mathbf{x}}^{\dagger}\left(\partial_{x_j} U_{\mathbf{x}}\right).
 \label{eq:generator}
\end{sequation}\noindent
Then, the quantum Fisher information matrix in \rSupeq{eq:qfim_pure} can be expressed in terms of the generators as
\begin{sequation}
	[F_Q]_{x_j x_k} = 2 \langle \psi_0 | \{ h_{x_j},  h_{x_k}\} | \psi_0\rangle - 4 \langle \psi_0 | h_{x_j} |\psi_0\rangle \langle \psi_0| h_{x_k} |\psi_0\rangle.
	\label{eq:generator_qfi}
\end{sequation}\noindent
Here $\{\cdot,\cdot\}$ denotes the anti-commutator. According to QCRB, the precision of estimating multiple parameters is lower bound by $n \text{Tr}(\text{Cov}(\mathbf{x}_{\text {est}}))\geq \text{Tr}(F_C^{-1})\geq  \text{Tr}(F_Q^{-1})$. Hence, finding the maximal QFIM and the optimal measurement that saturates the QCRB leads to the ultimate precision of estimation.
For multi-parameter quantum estimation, the necessary and sufficient condition for saturating the quantum Cramér-Rao bound in pure states is the weak commutativity condition, which is
\begin{sequation}
	\operatorname{Im}\left[\langle \partial_{x_j} \psi_{\mathbf{x}} | \partial_{x_k} \psi_{\mathbf{x}}\rangle\right] = 0, \forall x_j, x_k.
 \label{eq:weak}
\end{sequation}\noindent

To achieve the best precision for estimating unknown parameters $\mathbf{x}$,  it is crucial to optimize every step of the process: the initial state, the controls during the evolution, and the final measurement. In the control-enhanced sequential scheme, the total system evolution is described by $U_N = (U_c U_{\mathbf{x}})^{N}$, where $U_{\mathbf{x}}$ represents the system dynamics over time $T$, and $U_c$ is a control operation applied after each cycle. The optimal strategy, as derived in \cite{Yuan2016}, employs the control $U_c = U_{\mathbf{x}}^{\dagger}$. This choice leads to a quadratic enhancement in precision, scaling the quantum Fisher information matrix as $N^2 F_Q$, where $F_Q$ is the QFIM for a single cycle ($N=1$). 
\textcolor{black}{In the following analysis, we will first obtain the maximal $F_Q$ for a single cycle. The corresponding QFIM for a multi-cycle strategy with optimal control is then obtained by scaling $F_Q$ by $N^2$. We note that the maximal $F_Q$ for a single cycle is equivalent to the maximal QFIM achievable under the dynamics $U_{\mathbf{x}}$ without the control. This is because a single control operation applied only at the end of the evolution cannot improve the QFIM.}

\subsection{Sensing of a remote vector field}

We consider the estimation of three components of a remote vector field, described in spherical coordinates $(B, \theta, \phi)$ as $\mathbf{B} = (B\sin \theta \cos \phi, B \sin \theta \sin \phi, B \cos \theta)$, instead of in Cartesian coordinates $\mathbf{B} = (B_x, B_y, B_z)$. Estimating the vector field components $\mathbf{B} = (B_{x}, B_{y}, B_{z})$ thus corresponds to  simultaneously estimating the parameters $\mathbf{x} = (B,\theta, \phi)$. For each sensor qubit at time $T$, the evolution can be represented by $U_s = e^{-i  \mathbf{B} \cdot \boldsymbol{\sigma} T}  = e^{-i B T \boldsymbol{n} \cdot \boldsymbol{\sigma}}$ with $\boldsymbol{n} = (\sin \theta \cos \phi, \sin \theta \sin \phi, \cos \theta)$. The generator for $x_j \in \{B,\theta, \phi\}$ is given by 
\begin{sequation}
	\begin{aligned}
		h_{B} &= c_{B} \boldsymbol{n}_{B} \cdot \boldsymbol{\sigma} \\
		h_{\theta} &= c_{\theta} \boldsymbol{n}_{\theta} \cdot \boldsymbol{\sigma} \\
		h_{\phi} &= c_{\phi} \boldsymbol{n}_{\phi} \cdot \boldsymbol{\sigma} \\
	\end{aligned}
	\label{eq:generators}
\end{sequation}\noindent
with
\begin{sequation}
	c_{B} = T,\quad  c_{\theta} = \sin(B T),\quad c_{\phi} = \sin(B T) \sin\theta,
\end{sequation}\noindent
\begin{sequation}
	\begin{aligned}
		&\boldsymbol{n}_{B} = \boldsymbol{n} = (\sin\theta \cos\phi, \sin\theta \sin\phi, \cos\theta),\\
		&\boldsymbol{n}_{\theta} =  \cos (B T) \boldsymbol{n}_{1} - \sin (B T) \boldsymbol{n}_{2}, \\
		&\boldsymbol{n}_{\phi} =  \sin (B T) \boldsymbol{n}_{1} + \cos (B T) \boldsymbol{n}_{2},
	\end{aligned}
	\label{eq:n_xj}
\end{sequation}\noindent
where $\boldsymbol{n}_{1} = \partial_{\theta} \boldsymbol{n} = (\cos \theta \cos \phi, \cos \theta \sin \phi, -\sin \theta)$, $\boldsymbol{n}_{2} = \boldsymbol{n} \times \boldsymbol{n}_{1} = (-\sin \phi , \cos \phi, 0)$. It is easy to verify that $\boldsymbol{n}$, $\boldsymbol{n}_{1}$, $\boldsymbol{n}_{2}$ are orthogonal to each other, since there exists a unitary transformation $U_r=e^{i \frac{B T}{2} \boldsymbol{n} \cdot \boldsymbol{\sigma}} e^{-i \frac{\phi}{2} \sigma_z} e^{-i \frac{\theta}{2} \sigma_y}$ such that 
\begin{sequation}
	\boldsymbol{n}_{B} \cdot \boldsymbol{\sigma} = U_r \sigma_z U_r^{\dagger}, \quad \boldsymbol{n}_{\theta} \cdot \boldsymbol{\sigma} = U_r \sigma_x U_r^{\dagger}, \quad \boldsymbol{n}_{\phi} \cdot \boldsymbol{\sigma} = U_r \sigma_y U_r^{\dagger}.
	\label{eq:rotation}
\end{sequation}\noindent
Assume the initial probe state is $|\psi_{SA}\rangle$, where the ancilla system is introduced. Then using \rSupeq{eq:generator_qfi}, we obtain the QFIM as
\begin{sequation}
	F_Q = 4\left(\begin{matrix}
		T^2 - \left(\operatorname{Tr}(\rho_S h_{B})\right)^2 & - \operatorname{Tr}(\rho_S h_{B})\operatorname{Tr}(\rho_S h_{\theta}) &  -\operatorname{Tr}(\rho_S h_{B}) \operatorname{Tr}(\rho_S h_{\phi})\\
		- \operatorname{Tr}(\rho_S h_{B})\operatorname{Tr}(\rho_S h_{\theta}) & \sin^2(B T) - \left(\operatorname{Tr}(\rho_S h_{\theta})\right)^2 &  - \operatorname{Tr}(\rho_S h_{\theta})\operatorname{Tr}(\rho_S h_{\phi})\\
		- \operatorname{Tr}(\rho_S h_{B})\operatorname{Tr}(\rho_S h_{\phi}) &- \operatorname{Tr}(\rho_S h_{\theta})\operatorname{Tr}(\rho_S h_{\phi}) &  \sin^2(B T) \sin^2\theta - \left(\operatorname{Tr}(\rho_S h_{\theta})\right)^2
	\end{matrix}\right)
\end{sequation}\noindent
where $\rho_S = \operatorname{Tr}_A(|\psi_{SA}\rangle\langle\psi_{SA}|)$ denotes the reduced state by tracing out the ancilla system.
The maximal QFIM, denoted as $F_Q^{\max}$, is achieved when $\rho_S=\frac{1}{2}I$ with
\begin{sequation}
	F_Q^{\max} = 4\left(\begin{matrix}
		T^2  & 0 &  0\\
		0 & \sin^2(B T) & 0\\
		0 &0 &  \sin^2(B T)\sin^2 \theta
	\end{matrix}\right).
\end{sequation}\noindent
Here $F_Q^{\max}$ is maximal in the sense that $F_Q^{\max} -F_Q\geq 0$ for any other $F_Q$.
The optimal state can be chosen as any pure state $|\psi_{SA}\rangle$ with the reduced state $\rho_S = \frac{1}{2} I$. This can be any maximally entangled state, in particular, it can be chosen as $|\psi_{SA}\rangle = \frac{1}{\sqrt{2}}\left(|00\rangle + |11\rangle\right)$. 

Next we show the projective measurement in the Bell basis saturates the quantum Cramér-Rao bound. The projective measurement in the Bell basis is given by
\begin{sequation}
    M_{00} = \left|\Phi^{+}\right\rangle\left\langle\Phi^{+}\right|, \quad M_{01} = \left|\Phi^{-}\right\rangle\left\langle\Phi^{-}\right|, \quad M_{10} = \left|\Psi^{+}\right\rangle\left\langle\Psi^{+}\right|, \quad M_{11} = \left|\Psi^{-}\right\rangle\left\langle\Psi^{-}\right|,
\label{eq:bell_measurement}
\end{sequation}\noindent
where 
\begin{sequation}
\begin{aligned}
&\left|\Phi^{+}\right\rangle = \frac{1}{\sqrt{2}}\left(|00\rangle + |11\rangle\right), \quad \left|\Phi^{-}\right\rangle = \frac{1}{\sqrt{2}}\left(|00\rangle - |11\rangle\right), \\
&\left|\Psi^{+}\right\rangle = \frac{1}{\sqrt{2}}\left(|01\rangle + |10\rangle\right), \quad \left|\Psi^{-}\right\rangle = \frac{1}{\sqrt{2}}\left(|01\rangle - |10\rangle\right),
\label{eq:bell_basis}
\end{aligned}
\end{sequation}\noindent
are the Bell states. Under this measurement, the probabilities of the measurement outcomes are given by
\begin{sequation}
	\begin{aligned}
		&P_{00} = \operatorname{Tr}(\rho(B,\theta,\phi) M_{00}) = \cos ^2(B T), \\
		&P_{01} = \operatorname{Tr}(\rho(B,\theta,\phi) M_{01}) =\sin ^2(B T) \cos ^2 \theta, \\
		&P_{10} = \operatorname{Tr}(\rho(B,\theta,\phi) M_{10}) =\sin ^2(B T) \sin ^2 \theta \cos ^2 \phi, \\
		&P_{11} = \operatorname{Tr}(\rho(B,\theta,\phi) M_{11}) =\sin ^2(B T) \sin ^2 \theta \sin ^2 \phi, \\
	\end{aligned}
\end{sequation}\noindent
where $\rho(B,\theta,\phi) = (U_{s}\otimes I) |\psi_{SA}\rangle \langle \psi_{SA}| (U_{s}^{\dagger}\otimes I) $ is the evolved state. \textcolor{black}{From these probabilities, we can obtain the classical Fisher information matrix via \rSupeq{eq:cfim} as
\begin{sequation}
    F_C = 4\left(\begin{matrix}
		T^2  & 0 &  0\\
		0 & \sin^2(B T) & 0\\
		0 &0 &  \sin^2(B T)\sin^2 \theta
	\end{matrix}\right).
\end{sequation}\noindent
This is identical to the quantum Fisher information matrix $F_Q^{\max}$, which verifies the projective measurement in the Bell basis saturates the quantum Cramér-Rao bound.} The quantum Cramér-Rao bound $\text{Cov}(\mathbf{x}_{\text {est}}) \geq F_Q^{-1}$(here we neglect the repetition $n$, which is a classical effect and the same for all schemes)  is thus achievable, which leads to the precision for the three parameters in the spherical coordinates as\begin{sequation}
	\delta B_{{\text{est}}}^2 \geq \frac{1}{4T^2}, \quad \delta \theta_{{\text{est}}}^2 \geq \frac{1}{4\sin^2(BT)}, \quad\delta \phi_{{\text{est}}}^2 \geq \frac{1}{4\sin^2(BT) \sin^2 \theta}.
\end{sequation}\noindent
In the Euclidean coordinates, this gives an achievable bound on the total variances as
\begin{sequation}
	\begin{aligned}
		\delta B_{x_{\text{est}}}^2 + \delta B_{y_{\text{est}}}^2 + \delta B_{z_{\text{est}}}^2 &= \delta B_{{\text{est}}}^2 + B^2 \delta \theta_{{\text{est}}}^2 + B^2 \sin^2 \theta \delta \phi_{{\text{est}}}^2 \\
		&\geq\frac{1}{4T^2} +\frac{B^2}{2 \sin^2(B T)}.
	\end{aligned}
	\label{eq:propagation_3}
\end{sequation}\noindent

For $N$ cycles, the maximal quantum Fisher information matrix is given by $N^2F_Q^{\max}$. The corresponding precisions are 
\begin{sequation}
	\delta B_{{\text{est}}}^2 \geq \frac{1}{4N^2T^2}, \quad \delta \theta_{{\text{est}}}^2 \geq \frac{1}{4N^2\sin^2(BT)}, \quad\delta \phi_{{\text{est}}}^2 \geq \frac{1}{4N^2\sin^2(BT) \sin^2 \theta}.
\end{sequation}\noindent 

\textcolor{black}{In comparison, the individual strategy, referred to as the "classical individual measurement" protocol in~\cite{Hou2021}, involves measuring the system immediately after one cycle. The process is then repeated $N/3$ times for estimating one parameter---$N$ operators total for estimating all three parameters. The resulting precisions are given by:
    \[\delta B_{\text{est}}^2 \geq \frac{3}{4N T^2}, \quad \delta \theta_{\text{est}}^2 \geq \frac{3}{4 N \sin^2(B T)}, \quad \delta \phi_{\text{est}}^2 \geq \frac{3}{4N \sin^2(B T) \sin^2 \theta},\]
    which are taken as the standard quantum limit. }

\textcolor{black}{When the $N$ operators act sequentially on the system qubit  without the control, the total evolution is given by $[U_s(\mathbf{x})]^{N} = e^{-i N \mathbf{B}\cdot \boldsymbol{\sigma} T}$, here $\mathbf{x} = (B, \theta, \phi)$ and $\mathbf{B} = (B \sin \theta \cos \phi, B \sin \theta \sin \phi, B \cos \theta)$. 
In this case the maximal quantum Fisher information matrix is  
\begin{sequation}
	F_Q^{\max} = 4\left(\begin{matrix}
		N^2T^2  & 0 &  0\\
		0 & \sin^2(NB T) & 0\\
		0 &0 &  \sin^2(NB T)\sin^2 \theta
	\end{matrix}\right),
\end{sequation}\noindent
which gives
\begin{sequation}
\delta B_{\text{est}}^2 \geq \frac{1}{4 N^2 T^2}, \quad \delta \theta_{\text{est}}^2 \geq \frac{1}{4 \sin^2(N B T)}, \quad \delta \phi_{\text{est}}^2 \geq \frac{1}{4 \sin^2(N B T) \sin^2 \theta}.
\end{sequation}\noindent
While this strategy improves the scaling for the estimation of the amplitude $B$, its performance for the estimation of $\theta$ and $\phi$ can degrade significantly and even become worse than the individual strategy. In particular, under the experimental condition used in our study, $B T = 3\pi/2$, the sine terms vanish when $N$ is an even number, leading to diverging estimation errors for both $\theta$ and $\phi$. For this reason, we do not include this strategy for the comparison in the main text.}

\subsection{Sensing of the gradients between vector fields}

In this section, we consider the estimation of the gradients between two remote vector fields, expressed as $\nabla \mathbf{B}=\mathbf{B}_1-\mathbf{B}_2 = (\nabla B_x, \nabla B_y, \nabla B_z)$. 
For each sensor qubit, the Hamiltonian $H_j$ is given by $H_j = \mathbf{B}_j \cdot \boldsymbol{\sigma} = B_{jx} \sigma_x + B_{jy} \sigma_y +B_{jz} \sigma_z$, where $\mathbf{B}_j = (B_{jx}, B_{jy}, B_{jz})$ represents the vector field components and $\boldsymbol{\sigma} = (\sigma_x, \sigma_y, \sigma_z)$ denotes the spin vector.
Alternatively, in spherical coordinates, the Hamiltonian can be represented as $H_j= B_j \boldsymbol{n}_j\cdot \boldsymbol{\sigma}$, where $\boldsymbol{n}_j = (\sin\theta_j \cos\phi_j, \sin\theta_j \sin\phi_j, \cos\theta_j)$ and $B_j = \sqrt{B_{jx}^2 + B_{jy}^2 + B_{jz}^2}$ represents the magnitude of the vector field. 
The evolution at time $T$ can be represented by $U_{sj} = e^{-i  \mathbf{B}_j\cdot \boldsymbol{\sigma} T} = e^{-i B_j \boldsymbol{n}_j\cdot \boldsymbol{\sigma} T}$ for $j=1,2$. 

We compare the precision of two strategies for estimating a vector field's gradient. The first uses non-local entanglement to directly measure the gradient across the sensor network. The second employs only local entanglement, measuring the field independently at each point before calculating the difference. Our analysis confirms that non-local entanglement provides a distinct quantum advantage, offering superior precision for distributed gradient sensing.

\subsubsection{Strategy with non-local entanglement}
In this approach, we employ a 4-qubit non-local entangled state, $|\Psi_0\rangle = \frac{1}{\sqrt{2}} (|0011\rangle - |1100\rangle)$, as the probe state to directly estimate gradients along three directions. The first two and last two qubits are distributed to two separate sensor modules (denoted $\mathcal{B}$ and $\mathcal{C}$ in the main text) respectively. The system undergoes a total evolution described by $U_{\rm{S}} = U_{s1} \otimes U_{s1} \otimes U_{s2} \otimes U_{s2}$, which imparts information about the spatial gradients onto the evolved state $U_{\rm{S}}|\Psi_0\rangle$.
The components of the vector fields at the two locations can be expressed in terms of their sum and gradient:
\begin{sequation}
B_{1j} = \frac{\sum B_j + \nabla B_j}{2}, \quad B_{2j} = \frac{\sum B_j - \nabla B_j}{2}, \quad \text{for } j \in {x, y, z},
\end{sequation}
where $\sum\! \mathbf{B} = \mathbf{B}_1 + \mathbf{B}_2 = (\sum B_x, \sum B_y, \sum B_z)$ and $\nabla \mathbf{B}=\mathbf{B}_1 - \mathbf{B}_2 = (\nabla B_x, \nabla B_y, \nabla B_z)$.

To effectively estimate the gradients, it is also necessary to acquire information about the sum of the vector fields. This information is crucial for adaptively implementing control strategies. Such control strategies essentially reduce the problem of estimating general gradients $\nabla \mathbf{B}$ to estimating $\nabla \mathbf{B} = (0,0,0)$, indicating $\mathbf{B}_1 = \mathbf{B}_2 = \mathbf{B}$. This can be realized by adding compensation to the vector fields if the gradients are non-zero. Therefore, we benchmark the performance of our estimation protocol at zero-gradient, and the comparison of different strategies will also be made under this assumption.

The quantum Fisher information matrix for the simultaneous estimation of the parameters $\mathbf{x} = (\nabla \mathbf{B}, \sum\! \mathbf{B})$ under the dynamics $U_{\rm{S}}$ is block-diagonal:
\begin{sequation}
F_Q = \begin{pmatrix}
F_{-} & \boldsymbol{0} \\
\boldsymbol{0} & F_{+}
\end{pmatrix},
\label{eq:qfim_gradient_summation}
\end{sequation}
where $F_{-}$ and $F_{+}$ are the QFIM submatrices for estimating the gradient $\nabla \mathbf{B} = (\nabla B_x, \nabla B_y, \nabla B_z)$ and the sum $\sum\! \mathbf{B} = (\sum B_x, \sum B_y, \sum B_z)$, respectively. The matrix $F_{-}$ is given by
\begin{sequation}
    F_{-} = \left(\begin{matrix}
        [F_Q]_{\nabla B_x \nabla B_x} & [F_Q]_{\nabla B_x \nabla B_y} & [F_Q]_{\nabla B_x \nabla B_z} \\
        [F_Q]_{\nabla B_x \nabla B_y} & [F_Q]_{\nabla B_y \nabla B_y} & [F_Q]_{\nabla B_y \nabla B_z}\\
        [F_Q]_{\nabla B_x \nabla B_z} & [F_Q]_{\nabla B_y \nabla B_z} & [F_Q]_{\nabla B_z \nabla B_z}
    \end{matrix}\right),
\end{sequation}\noindent
with its elements 
\begin{equation}\footnotesize
    \begin{aligned}
        [F_Q]_{\nabla B_x \nabla B_x} =& \frac{4}{B^4} \left(B_x^2 T^2 (B^2 + 3 B_z ^2) +\sin^2(B T) (B_y^2 + B_z^2 + 6 B_x B_y B_z T + 3 B_y^2 \sin^2(BT))\right)\\
        &+\frac{3 B_x B_z \sin(2 B T)}{B^6} \left( - 4 B B_y \sin^2(B T) + B_x B_z (-4BT + \sin(2 B T) \right),\\
        [F_Q]_{\nabla B_y \nabla B_y} =&  \frac{4}{B^4} \left( B_y^2 T^2 (B^2 + 3 B_z ^2) +\sin^2(B T) ( B_x^2 + B_z^2 - 6 B_x B_y B_z T + 3 B_x^2 \sin^2 (BT))\right)\\
        &+\frac{3 B_y B_z \sin(2 B T)}{B^6} \left( 4 B B_x \sin^2(B T) + B_y B_z (-4BT + \sin(2 B T)\right)\\
        [F_Q]_{\nabla B_z \nabla B_z} =& \frac{4}{B^4} \left( B_z^2 T^2 (B^2 + 3 B_z ^2) + \sin^2(B T) (B_x^2 + B_y^2)\right)\\
        &+\frac{3 \sin(2 B T)}{B^6} \left( 4 B (B_x^2 +B_y^2 )B_z^2 T + (B_x^2 +B_y^2 )^2 \sin(2 B T) \right),  \\
        [F_Q]_{\nabla B_x \nabla B_y} =&  \frac{4}{B^4} \left( B_x B_y T^2 (B^2 + 3 B_z ^2) - \sin^2(B T) (4 B_x B_y + 3 B_z T(B_x^2 -B_y^2) )\right)\\
        &+\frac{3 \sin(2 B T)}{B^6} \left( 2 B B_z \sin^2(B T)  (B_x^2 -B_y^2 ) + B_x B_y (-4 B B_z^2 T + (B^2 + B_z^2)\sin(2 B T)) \right),  \\
        [F_Q]_{\nabla B_x \nabla B_z} =&  \frac{4 B_z}{B^4} \left( B_x T^2 (B^2 + 3 B_z ^2) - \sin^2(B T) ( B_x- 3 B_y B_z T )\right)\\
        &+\frac{3 \sin(2 B T)}{B^6} \left( 2 B B_y \sin^2(B T)  (B_x^2 +B_y^2 ) + B_x B_z (2 B T (B^2 - 2 B_z^2) - (B^2 - B_z^2)\sin(2 B T)) \right),  \\
        [F_Q]_{\nabla B_y \nabla B_z} =& \frac{4 B_z}{B^4} \left( B_y T^2 (B^2 + 3 B_z ^2) - \sin^2(B T) ( B_y + 3 B_x B_z T)\right)\\
        &-\frac{3 \sin(2 B T)}{B^6} \left( 2 B B_x \sin^2(B T)  (B_x^2 + B_y^2 ) - B_y B_z (2 B T (B^2 - 2 B_z^2) - (B^2 - B_z^2)\sin(2 B T)) \right).  \\
    \end{aligned}
\end{equation}
Similarly, the matrix $F_{+}$ is given by
\begin{sequation}
    F_{+} = \left(\begin{matrix}
        [F_Q]_{\sum B_x \sum B_x} & [F_Q]_{\sum B_x \sum B_y} & [F_Q]_{\sum B_x \sum B_z} \\
        [F_Q]_{\sum B_x \sum B_y} & [F_Q]_{\sum B_y \sum B_y} & [F_Q]_{\sum B_y \sum B_z}\\
        [F_Q]_{\sum B_x \sum B_z} & [F_Q]_{\sum B_y \sum B_z} & [F_Q]_{\sum B_z \sum B_z}
    \end{matrix}\right),
\end{sequation}\noindent
with its elements
\begin{equation}\footnotesize
    \begin{aligned}
        [F_Q]_{\sum B_x \sum B_x} =& \frac{4}{B^4} \left(B_x^2 T^2 (B_x^2 + B_y ^2) + B_z \sin^2(B T) (B_z - 2 B_x B_y T)\right)\\
        &+\frac{\sin(2 B T)}{B^6} \left( 4 B B_x^2 B_z^2 T +4 B B_x B_y B_z \sin^2(B T)+ (B^2 B_y^2 -B_x^2 B_z^2) \sin(2 B T) \right),\\
        [F_Q]_{\sum B_y \sum B_y} =& \frac{4}{B^4} \left(B_y^2 T^2 (B_x^2 + B_y ^2) + B_z \sin^2(B T) (B_z + 2 B_x B_y T)\right)\\
        &+\frac{\sin(2 B T)}{B^6} \left( 4 B B_y^2 B_z^2 T -4 B B_x B_y B_z \sin^2(B T)+ (B^2 B_x^2 -B_y^2 B_z^2) \sin(2 B T) \right),\\
        [F_Q]_{\sum B_z \sum B_z} =& \frac{4}{B^4} (B_x^2 + B_y^2)(B_z^2 T^2 + \sin^2 (B T) )\\
        &+\frac{\sin(2 B T)}{B^6} (B_x^2 + B_y^2)\left(4 B B_z^2 T + (B_x^2 + B_y^2) \sin(2 B T) \right)\\
        [F_Q]_{\sum B_x \sum B_y} =& \frac{4 T}{B^4} \left( B_x B_y T (B_x^2 + B_y^2) +  B_z \sin^2 (BT)(B_x^2 - B_y^2) \right) \\
        &+ \frac{\sin(2B T)}{B^6}\left(4 B B_x B_y B_z^2 T - 2 B B_z \sin^2(B T )(B_x^2 - B_y^2) - B_x B_y \sin(2 B T) (B^2 + B_z^2) \right),  \\
        [F_Q]_{\sum B_x \sum B_z} =& \frac{4 B_z}{B^4} \left( B_x T^2 (B_x^2 + B_y^2) -  \sin^2 (BT)(B_x+ B_y B_z T) \right) \\
        &- \frac{\sin(2B T)}{B^6}\left(2 B B_x B_z T(B^2 - 2B_z^2) + (B_x^2 + B_y^2) \left( 2 B B_y \sin^2 (B T) - B_x B_z \sin(2 B T) \right) \right),  \\
        [F_Q]_{\sum B_y \sum B_z} =& \frac{4 B_z}{B^4} \left( B_y T^2 (B_x^2 + B_y^2) -  \sin^2 (BT)(B_y - B_x B_z T) \right) \\
        &- \frac{\sin(2B T)}{B^6}\left(2 B B_y B_z T(B^2 - 2B_z^2) - (B_x^2 + B_y^2) \left( 2 B B_x \sin^2 (B T) + B_y B_z \sin(2 B T) \right) \right).  \\
    \end{aligned}
\end{equation}
Due to the block-diagonal structure of the quantum Fisher information matrix, the precision of estimating the gradient components $\nabla \mathbf{B}$ is independent of the estimation of the sum field $\sum\! \mathbf{B}$. The total variance for estimating the gradient is bounded by:
\begin{sequation}
\left(\delta \nabla B_{x_{\text{est}}}\right)^2+\left(\delta \nabla B_{y_{\text{est}}}\right)^2+\left(\delta \nabla B_{z_{\text{est}}}\right)^2 \geq \operatorname{Tr}\left(F_{-}^{-1}\right)= \frac{4 B^2-3 B_z^2}{16 B^2 T^2}+\frac{5 B^2+3 B_z^2}{16 \sin ^2 (B T)},
\label{eq:pre_nle_3}
\end{sequation}
where $B = \sqrt{B_x^2 + B_y^2 + B_z^2}$. We have verified that the weak commutativity condition (\rSupeq{eq:weak}) is satisfied, confirming that a set of POVMs exists which can saturate the quantum Cramér-Rao bound and achieve this precision.

For measurement, we consider local separable measurements on each sensor module ($\mathcal{B}$ and $\mathcal{C}$). Specifically, we choose a projective measurement in the Bell basis for each module:
\begin{sequation}
\Pi_{\mathcal{X}_{00}} = \left|\Phi^{+}\right\rangle\left\langle\Phi^{+}\right|, \quad \Pi_{\mathcal{X}_{01}} = \left|\Phi^{-}\right\rangle\left\langle\Phi^{-}\right|, \quad \Pi_{\mathcal{X}_{10}} = \left|\Psi^{+}\right\rangle\left\langle\Psi^{+}\right|, \quad \Pi_{\mathcal{X}_{11}} = \left|\Psi^{-}\right\rangle\left\langle\Psi^{-}\right|,
\label{eq:bell_measurement_local}
\end{sequation}\noindent
where $\mathcal{X}\in \{\mathcal{B}, \mathcal{C}\}$ denoted the sensor module and the Bell states are defined in \rSupeq{eq:bell_basis}.
This defines 16 measurement operators $\Pi_k= \Pi_{\mathcal{B}_i} \otimes \Pi_{\mathcal{C}_j}$ for $i,j \in \{00,01,10,11\}$, which form a complete set ($\sum_k \Pi_k = I$). 

The probability of outcome $k$ is $P_k =  \langle \Psi_{\mathbf{x}}| \Pi_k |\Psi_{\mathbf{x}}\rangle$ for $k \in \{0000, 0001,\cdots, 1111\}$, where $|\Psi_{\mathbf{x}}\rangle = U_{\rm{S}}|\Psi_0\rangle$ is the evolved state. For outcomes in the set $\{0011, 0111, 1011, 1100, 1101, 1110, 1111\}$, $P_k = 0$ for all parameter values, providing no information. For outcomes in $\{0000, 0101, 1010\}$, $P_k = 0$ specifically at the zero-gradient point ($\nabla \mathbf{B} = (0,0,0)$). Despite these zeros, the probability distribution still encodes information about the parameters.

The classical Fisher information matrix (CFIM) depends on the derivative of $P_k$ with respect to $x_i, x_j \in \{\nabla B_x, \nabla B_y, \nabla B_z, \sum B_x, \sum B_y, \sum B_z\}$,
\begin{sequation}
  \begin{aligned}
    [F_C]_{x_i x_j} &= \sum_{k} \frac{1}{P_k}\left(\frac{\partial P_k}{\partial x_i
    }\right)\left(\frac{\partial P_k}{\partial x_j}\right) \\
    & = \sum_{k} \frac{\left(\partial_{x_i} \langle \Psi_{\mathbf{x}}| \Pi_k | \Psi_{\mathbf{x}}\rangle \right)\left(\partial_{x_j} \langle \Psi_{\mathbf{x}}| \Pi_k | \Psi_{\mathbf{x}}\rangle \right)}{\langle \Psi_{\mathbf{x}}| \Pi_k | \Psi_{\mathbf{x}}\rangle} \\
    & = \sum_{k} \frac{4 \mathrm{Re} \left( \langle \partial_{x_i} \Psi_{\mathbf{x}}| \Pi_k | \Psi_{\mathbf{x}}\rangle \right)\mathrm{Re} \left(\langle \partial_{x_j}  \Psi_{\mathbf{x}}| \Pi_k | \Psi_{\mathbf{x}}\rangle \right)}{\langle\Psi_{\mathbf{x}}| \Pi_k |\Psi_{\mathbf{x}} \rangle},
  \end{aligned}
\end{sequation}\noindent
where the sum is over the informative outcomes $k \in \{0000, 0001, 0010,0100, 0101,0110,1000, 1001, 1010\}$. At the zero-gradient point, $\nabla B_x = 0,\nabla B_y = 0$ and $\nabla B_z = 0$, $P_k = \langle \Psi_{\mathbf{x}}| \Pi_k | \Psi_{\mathbf{x}}\rangle  = 0$, $\mathrm{Re} \left( \langle \partial_{x_i} \Psi_{\mathbf{x}}| \Pi_k | \Psi_{\mathbf{x}}\rangle \right) = 0$ for $k \in \{0000, 0101, 1010\}$. 
In these cases, the term is of the form $\frac{0}{0}$ that needs to be calculated through the limit by considering an infinitesimal displacement of $\nabla B_x, \nabla B_y$ and $\nabla B_z$, replacing $| \Psi_{\mathbf{x}}\rangle$ with $| \Psi_{\mathbf{x}}\rangle +\sum_{l=1}^{6} \delta x_{l}\left|\partial_{x_{l}}\Psi_{\mathbf{x}}\right\rangle$. 
It can then be verified that for all parameters $x_i, x_j$, $\forall x_i, x_j \in\{\nabla B_x, \nabla B_y, \nabla B_z, \sum B_x, \sum B_y, \sum B_z\}$,
\begin{sequation}
  \begin{aligned}
    [F_C]_{x_i x_j} &= \sum_{k_1}\frac{\sum_{l_1=1}^6 \sum_{l_2=1}^6 4 \delta x_{l_1} \delta x_{l_2} \mathrm{Re} \left( \langle \partial_{x_i} \Psi_{\mathbf{x}}| \Pi_{k_1} |\partial_{x_{l_1}} \Psi_{\mathbf{x}}\rangle \right)\mathrm{Re} \left(\langle \partial_{x_{l_2}}  \Psi_{\mathbf{x}}| \Pi_{k_1} | \partial_{x_j} \Psi_{\mathbf{x}}\rangle \right)}{\sum_{l_1=1}^6 \sum_{l_2=1}^6  \delta x_{l_1} \delta x_{l_2} \langle \partial_{x_{l_1}} \Psi_{\mathbf{x}}| \Pi_{k_1} |\partial_{x_{l_2}} \Psi_{\mathbf{x}}\rangle} \\
    & \quad + \sum_{k_2} \frac{4 \mathrm{Re} \left( \langle \partial_{x_i} \Psi_{\mathbf{x}}| \Pi_{k_2} | \Psi_{\mathbf{x}}\rangle \right)\mathrm{Re} \left(\langle \partial_{x_j}  \Psi_{\mathbf{x}}| \Pi_{k_2} | \Psi_{\mathbf{x}}\rangle \right)}{\langle \Psi_{\mathbf{x}}| \Pi_{k_2} | \Psi_{\mathbf{x}} \rangle} \\
    &= [F_Q]_{x_i x_j},
  \end{aligned}
\end{sequation}\noindent
where $k_1 \in \{0000, 0101, 1010\}$ and $k_2 \in \{0001,0010,0100,0110,1000,1001\}$. \textcolor{black}{The classical Fisher information matrix $F_C$ is thus equal to the quantum Fisher information matrix $F_Q$. This demonstrates that the QCRB can be saturated by performing local projective measurements in the Bell basis on each sensor module, proving that the precision in \rSupeq{eq:pre_nle_3} is achievable.}

We also analyze the precision for estimating the gradients of two-dimensional vector fields, namely $\nabla B_x$ and $\nabla B_y$, using the same non-local entangled probe state $|\Psi_0\rangle$. The QFIM for estimating ${\nabla B_x, \nabla B_y, \sum B_x, \sum B_y}$ remains block-diagonal as in \rSupeq{eq:qfim_gradient_summation}, with the submatrices given by:
\begin{sequation}
    F_- = \left(\begin{matrix}
        [F_Q]_{\nabla B_x \nabla B_x} & [F_Q]_{\nabla B_x \nabla B_y} \\
        [F_Q]_{\nabla B_x \nabla B_y} & [F_Q]_{\nabla B_y \nabla B_y}
    \end{matrix}\right), \quad F_+ = \left(\begin{matrix}
        [F_Q]_{\sum B_x \sum B_x} & [F_Q]_{\sum B_x \sum B_y} \\
        [F_Q]_{\sum B_x \sum B_y} & [F_Q]_{\sum B_y \sum B_y}
    \end{matrix}\right),
\end{sequation}\noindent
where
\begin{sequation}
    \begin{aligned}
        [F_Q]_{\nabla B_x \nabla B_x} = &  \frac{4B_x^2 T^2}{B^2}+ \frac{B_y^2 \left(16 \sin^2( B T) -3 \sin^2(2 B T)\right)}{B^4}, \\
        [F_Q]_{\nabla B_y \nabla B_y} = &  \frac{4B_y^2 T^2}{B^2}+ \frac{B_x^2 \left(16 \sin^2( B T) -3 \sin^2(2 B T)\right)}{B^4},\\
        [F_Q]_{\nabla B_x \nabla B_y} = &  \frac{4B_x B_y T^2}{B^2}- \frac{B_x B_y \left(16 \sin^2( B T) -3 \sin^2(2 B T)\right)}{B^4},\\
        [F_Q]_{\sum B_x \sum B_x} = & \frac{4 B_x^2 T^2}{B^2} + \frac{B_y^2 \sin^2(2 B T)}{B^4},\\
        [F_Q]_{\sum B_y \sum B_y} = & \frac{4 B_y^2 T^2}{B^2} + \frac{B_x^2 \sin^2(2 B T)}{B^4},\\
        [F_Q]_{\sum B_x \sum B_y} = &  \frac{4 B_x B_y T^2}{B^2} - \frac{B_x B_y \sin^2(2 B T)}{B^4},
    \end{aligned}
\end{sequation}\noindent
with $B = \sqrt{B_x^2 + B_y^2}$.
\textcolor{black}{The quantum Cramér-Rao bound can also be saturated by performing projective measurement in Bell basis on each sensor module. It can be similarly verified by computing the classical Fisher information matrix using \rSupeq{eq:cfim} and comparing it with the quantum Fisher information matrix. The calculation follows a similar procedure to the three-component estimation case and is omitted here for brevity.}
The total variance for estimating the in-plane gradients can be similarly obtained as
\begin{sequation}
	\begin{aligned}
		\delta \nabla B_{x_{\text {est }}}^2+\delta \nabla B_{y_{\text {est }}}^2 &\geq \frac{1}{4 T^2}+\frac{B^2}{4\left(1+3 \sin ^2 (B T)\right) \sin ^2 (B T)}.
	\end{aligned}
\end{sequation}\noindent

\paragraph{Adaptive scheme for non-zero gradients.}

\textcolor{black}{ In the experiment, the vector field gradients are measured under the condition $\mathbf{B}_1 = \mathbf{B}_2$, corresponding to zero gradient. However, this can be generalized to arbitrary, non-zero gradients by locally adding a compensation field $\nabla \mathbf{B}_{\text{est}}$ at one sensor node.  In each iteration this field is updated, reducing the effective problem to estimating the residual gradient $\nabla \mathbf{B} - \nabla \mathbf{B}_{\text{est}}$, which becomes vanishingly small with the increase of the iteration. Asymptotically, the estimation precision converges to the case with zero gradient. }

\begin{figure*}[hbt]
	\centering
	\includegraphics[width=0.85\textwidth]{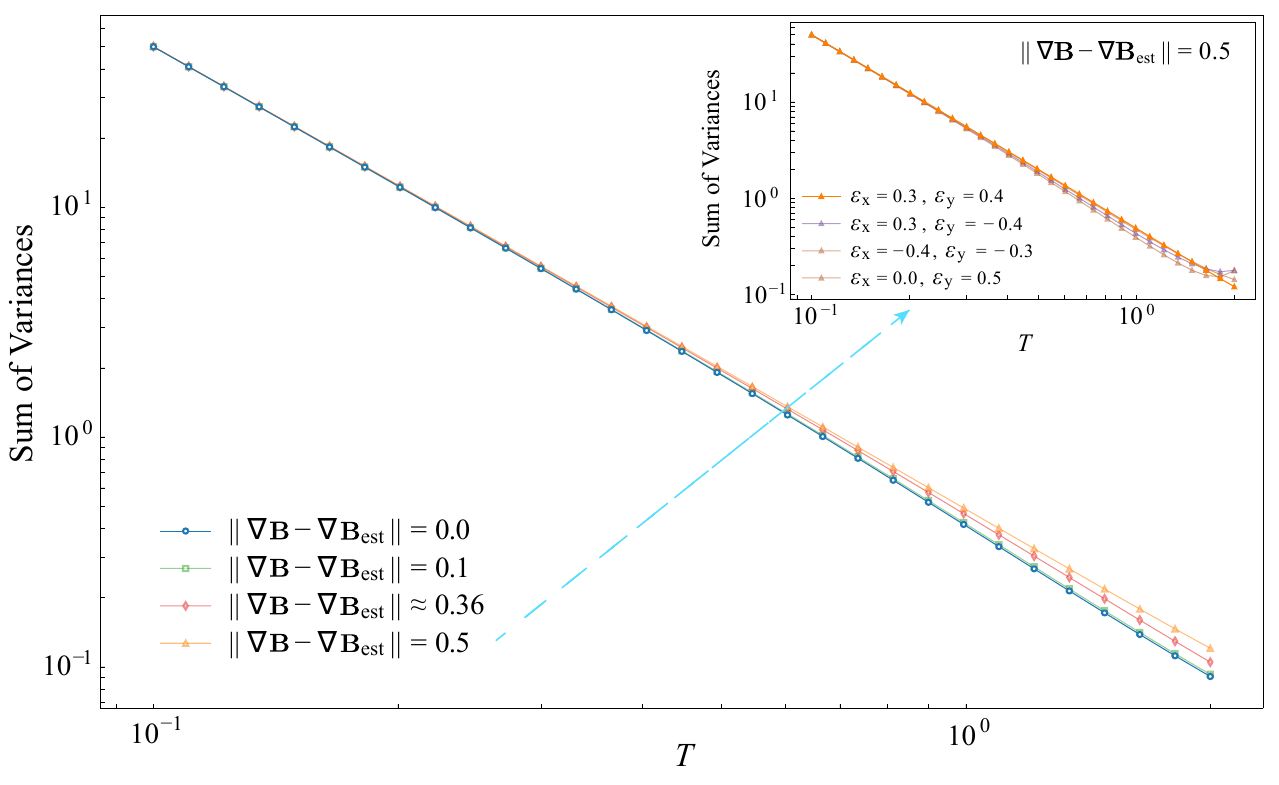}
	\caption{ 
        { {\bf Scaling of sum of variances under different residual gradient norms.} Main Panel: Log–log plot of the total estimation variance for the residual gradient $\nabla \mathbf{B} - \nabla \mathbf{B}_{\text{est}}$ as a function of total evolution time $T$. The true magnetic fields are set as $\mathbf{B}_1 = (\frac{\sqrt{2}}{4}, \frac{\sqrt{2}}{4}, 0)$ and $\mathbf{B}_2 = (0, 0, 0)$, yielding a true gradient $\nabla \mathbf{B} = (\frac{\sqrt{2}}{4}, \frac{\sqrt{2}}{4}, 0)$. A coarse estimate $\nabla \mathbf{B}_{\text{est}} = (B_{x\text{est}}, B_{y\text{est}}, 0)$ is applied at site 2, resulting in a residual gradient $\nabla \mathbf{B} - \nabla \mathbf{B}_{\text{est}} = (\epsilon_x, \epsilon_y, 0)$ to be estimated. The four curves correspond to residual norms of $\|\nabla \mathbf{B} - \nabla \mathbf{B}_{\text{est}}\| = 0 (\epsilon_x = \epsilon_y =0)$, $\|\nabla \mathbf{B} - \nabla \mathbf{B}_{\text{est}}\| = 0.1 (\epsilon_x = 0, \epsilon_y =0.1)$, $\|\nabla \mathbf{B} - \nabla \mathbf{B}_{\text{est}}\| \approx 0.36 (\epsilon_x =0.2, \epsilon_y =0.3)$, and $\|\nabla \mathbf{B} - \nabla \mathbf{B}_{\text{est}}\| = 0.5 (\epsilon_x =0.3, \epsilon_y =0.4)$, respectively.
        Inset: Scaling behavior of the total estimation variance for different orientations of the residual gradient, with the residual norm $\|\nabla \mathbf{B} - \nabla \mathbf{B}_{\text{est}}\|$ fixed at 0.5.}
        }
	\label{NLE_adap}
\end{figure*}

\rSupFig{NLE_adap} shows a numerical study of the achievable precision as a function of the residual gradient magnitude.  Even for offsets as large as $\|\nabla\mathbf{B}-\nabla\mathbf{B}_{\text{est}}\|=0.5$, the estimation precision remains close to the ideal case over a sufficiently long evolution time $T$. The inset in the figure further illustrates how deviations along the $x$ and $y$ directions affect the estimation performance. In practice, both the compensation field and the evolution time $T$ can be adaptively updated, allowing the protocol to converge asymptotically to the optimal precision. This strategy is conceptually similar to those employed in adaptive quantum metrology~\cite{pang2017optimal}, where coarse initial estimates are iteratively refined to achieve near-optimal scaling.

\subsubsection{Strategy with local entanglement}

To estimate the gradients between remote vector fields, a straightforward approach is to first estimate each vector field independently, then calculate the gradient. We first consider the estimation of two gradient components along the $x$ and $y$ directions, $(\nabla B_x, \nabla B_y)$. In this strategy, pairs of locally entangled two‑qubit states are used to estimate the local vector fields at two spatially separated sites; the gradient components are subsequently obtained by finite differencing of the two estimates. At each site, the two‑dimensional magnetic field is parameterized as  $\mathbf{x} = (B, \phi)$, from which the Cartesian components $B_x$ and $B_y$ can be directly inferred~\cite{hou2020}.

In the experiment, the initial probe state is prepared as the maximally entangled state $|\psi\rangle = \frac{1}{\sqrt{2}}\left(|00\rangle + |11\rangle\right)$. Each qubit undergoes the same local unitary evolution $U_s = e^{-i B T \boldsymbol{n} \cdot \boldsymbol{\sigma}}$, resulting in the two-qubit dynamics $U_s \otimes U_s$.
A projective measurement in the Bell basis (see \rSupeq{eq:bell_measurement}) is then performed on the evolved state, yielding the outcome probabilities:
\begin{sequation}
    \begin{aligned}
        P_{00} &= \left(\cos^2(B T) - \cos(2\phi) \sin^2 (B T)\right)^2,\\
        P_{01} &= \sin^4 (B T) \sin^2 (2\phi),\\
        P_{10} &= 4 \cos^2(B T) \cos^2 \phi \sin^2(B T),\\
        P_{11} &= 0.
    \end{aligned}
\end{sequation}\noindent
The classical Fisher information matrix for estimating $\mathbf{x} = \{B, \phi\}$ can then be obtained as
\begin{sequation}
    F_C = \left(\begin{matrix}
		16 T^2 \cos^2 \phi & -4 T\sin(2 B T )\sin(2\phi) \\
		-4 T \sin(2 B T)\sin(2\phi) & 7 - 8 \cos(2 B T) + 2 \cos(4 B T) \cos^2 \phi - \cos(2\phi)
	\end{matrix}\right).
\end{sequation}\noindent
This coincides with the QFIM, indicating the measurement saturates the quantum Cramér-Rao bound. Consequently, the variances of the estimators are bounded by
\begin{sequation}
	\begin{aligned}
		&\delta B_{\text {est }}^2 = [\text{Cov}(\mathbf{x}_{\text {est}})]_{11} \geq [F_C^{-1}]_{11} = \frac{1-\cos^2(B T)\cos^2 \phi}{16T^2\sin^2(B T)\cos^2 \phi}, \\
		&\delta \phi_{\text {est }}^2 = [\text{Cov}(\mathbf{x}_{\text {est}})]_{22} \geq [F_C^{-1}]_{22} = \frac{1}{16\sin^4(B T)}.
	\end{aligned}
\end{sequation}\noindent
The precision for estimating the two orthogonal components of the vector field using maximally entangled states is therefore given by
\begin{sequation}
		\delta B_{x_{\text {est }}}^2+\delta B_{y_{\text {est }}}^2  =\delta B_{\text {est }}^2+B^2 \delta \phi_{\text {est }}^2 \geq \frac{1}{16 \sin^2(B T)}\left(\frac{B^2-\cos^2(B T)B_x^2}{T^2 B_x^2} + \frac{B^2}{\sin^2(B T)}\right).
\end{sequation}\noindent
Gradients are estimated by differencing the independent estimates from two sites, which introduces an overall factor of two in the summed gradient precision:
\begin{sequation}
	\delta \nabla B_{x_{\text {est }}}^2+\delta \nabla B_{y_{\text {est }}}^2 \geq \frac{1}{8 \sin^2(B T)}\left(\frac{B^2-\cos^2(B T)B_x^2}{T^2 B_x^2} + \frac{B^2}{\sin^2(B T)}\right).
\end{sequation}\noindent

While the maximally entangled state is used experimentally for its ease of implementation, the precision within this local-entanglement strategy can be further improved by optimizing the input probe state. For a fair comparison, we now present the optimal probe and its corresponding theoretical lower bound, which follows the analysis in \cite{hou2020}. The optimal initial probe state in this case is given by 
\begin{sequation}
	|\psi_{opt}\rangle = \frac{1}{\sqrt{2}}(U_r \otimes U_r)(|11\rangle - |00\rangle),
    \label{eq:optimal_local}
\end{sequation}\noindent 
where $U_r=e^{i \frac{B T}{2} \boldsymbol{n} \cdot \boldsymbol{\sigma}} e^{-i \frac{\phi}{2} \sigma_z} e^{-i \frac{\pi}{4} \sigma_y}$. This optimal state depends on the true values of the unknown parameters $B$ and $\phi$. As they are not known a priori, an adaptive strategy needs to be adopted in practice by replacing the true values with their current estimates, $B_{\text{est}}$ and $\phi_{\text{est}}$. This yields the practical preparation unitary $U_r^{\text{est}}= e^{i \frac{B_{\text{est}} T}{2} \boldsymbol{n} \cdot \boldsymbol{\sigma}} e^{-i \frac{\phi_{\text{est}}}{2} \sigma_z} e^{-i \frac{\pi}{4} \sigma_y}$. In the asymptotical limit when the estimates converge to the true value, we obtain the highest precision as 
\begin{sequation}
		\delta B_{x_{\text {est }}}^2+\delta B_{y_{\text {est }}}^2  =\delta B_{\text {est }}^2+B^2 \delta \phi_{\text {est }}^2  \geq  \frac{1}{4} \left( \frac{1}{\langle \Delta^2 G_{B}\rangle } + \frac{B^2}{\langle \Delta^2 G_{\phi}\rangle } \right)
		\geq  \frac{1}{16 T^2}+\frac{B^2}{16 \sin ^2(B T)}.
	\label{eq:precision_analysis_2}
\end{sequation}\noindent
This bound is achievable by performing projective measurement in the Bell basis. Gradients are again estimated by differencing the independent estimates from two sites, which introduces an overall factor of two in the summed gradient precision:
\begin{sequation}
	\delta \nabla B_{x_{\text {est }}}^2+\delta \nabla B_{y_{\text {est }}}^2 \geq \frac{1}{8 T^2}+\frac{B^2}{8 \sin ^2(B T)}.
\end{sequation}\noindent

We then consider the estimation of gradients of three dimensional fields, $(\nabla B_x,\nabla B_y,\nabla B_z)$. Similarly, pairs of locally entangled two‑qubit states are used to estimate the local vector fields at two sites; the gradient components are subsequently obtained by finite differencing of the two estimates. At each site, the two‑dimensional magnetic field is parameterized as $\mathbf{x}=(B,\theta,\phi)$, from which the Cartesian components can be obtained.

To ensure a fair comparison, we again take the maximally entangled probe state $|\psi\rangle = \frac{1}{\sqrt{2}} (|00\rangle + |11\rangle)$. Each qubit undergoes the same local unitary evolution $U_s = e^{-i B T \boldsymbol{n} \cdot \boldsymbol{\sigma}}$, resulting in the two-qubit dynamics $U_s \otimes U_s$. The QFIM for estimating $\mathbf{x} = \{B,\theta,\phi\}$ is given by 
\begin{sequation}
    F_Q = \left(\begin{matrix}
        [F_Q]_{B B} & [F_Q]_{B \theta} & [F_Q]_{B \phi}\\
        [F_Q]_{B \theta} & [F_Q]_{\theta \theta} & [F_Q]_{\theta \phi}\\
        [F_Q]_{B \phi} & [F_Q]_{\theta \phi} & [F_Q]_{\phi \phi}\\
    \end{matrix}\right)
\end{sequation}\noindent
with
\begin{sequation}
    \begin{aligned}
        [F_Q]_{B B} =& 4 T^2\left(3+\cos (2\theta) + 2\cos (2\phi) \sin^2 \theta\right) ,\\ 
        [F_Q]_{\theta \theta}  =& 2\sin^2(B T) \left(3+ 3\cos(2 B T)\cos(2\phi) + 2\sin^2 \phi + \cos^2(B T) \left(2-4\cos(2\theta) \sin^2 \phi \right) \right. 
        \\&\left. + 4\cos \theta \sin(2 B T)\sin(2\phi)\right) ,\\
        [F_Q]_{\phi \phi} =& 2\sin^2(B T) \sin^2 \theta \left(2+ 2\sin^2 (B T)+ 2\sin^2 \theta + 2\sin^2 \phi +\cos(2 B T) \left(\cos(2\theta) -3 \cos(2\phi) \right)\right. \\&\left. + 2\sin^2(B T)\cos(2\theta)\cos(2\phi) - 4\sin(2BT)\cos\theta \sin(2\phi) \right),\\
        [F_Q]_{B \theta} = & 4 T\left( 2 \sin^2(BT) \sin \theta \sin(2\phi) - \sin(2 B T) \sin(2\theta) \sin^2 \phi\right),\\
        [F_Q]_{B \phi} = & -16 T \sin(B T) \sin^2 \theta \sin \phi \left( \cos(B T) \cos \phi + \sin(B T ) \cos \theta \sin \phi \right) ,\\
        [F_Q]_{\theta \phi} = & 2 \sin^2(B T) \left( \sin(2BT) \sin \theta \left( (3+\cos 2\theta)\cos(2\phi) + 2 \sin^2\theta\right) - 2\cos(2 B T) \sin(2\theta) \sin(2\phi) \right) .
    \end{aligned}
\end{sequation}\noindent
The resulting QFIM is singular, implying that, for the maximally entangled probe state considered here, the three parameters are not simultaneously identifiable. Consequently, within this local-entanglement strategy based on the maximally entangled state, a direct three-parameter benchmark analogous to the two-parameter case is not available.

\subsubsection{Additional strategy and comparison}
\begin{figure*}[hbt]
    \begin{center}
        \includegraphics[width=0.6\textwidth]{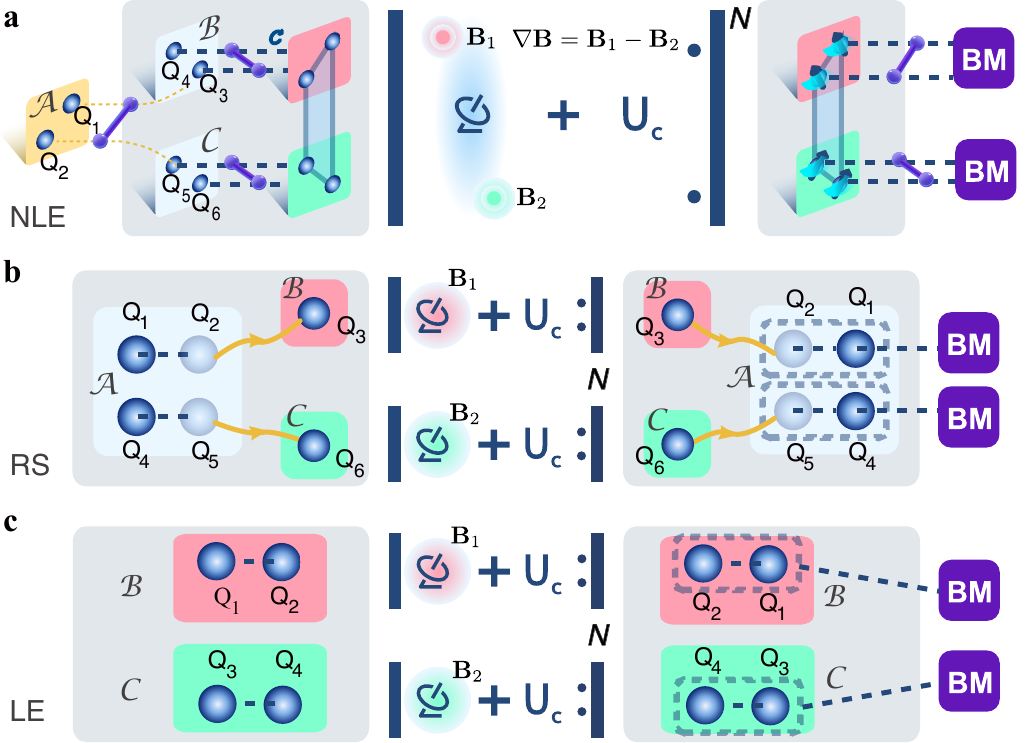}
        \caption{
            \label{separateStrategies}
            {\bf The sequences for measuring the gradient by NLE strategy, and separately measuring vector fields $\mathbf{B}_1$ and $\mathbf{B}_2$, and calculating their difference (RS/LE strategy).}
            {\bf a, } Non-local entanglement strategy with three modules, where $Q_3$, $Q_4$ on $\mathcal{B}$ and $Q_5$, $Q_6$ on $\mathcal{C}$ serve as the sensor qubits.
            {\bf b, } Remote sensing strategy with modules $\mathcal{A}-\mathcal{B}$ and $\mathcal{A}-\mathcal{C}$, where $Q_3$ on $\mathcal{B}$ and $Q_6$ on $\mathcal{C}$ act as the sensor qubits, $Q_1$ and $Q_4$ on $\mathcal{A}$ serve as the ancilla qubits.
            {\bf c, } Local entanglement strategy with modules $\mathcal{B}$ and $\mathcal{C}$, where $\mathbf{B}_1$ is applied on $Q_1$ and $Q_2$, $\mathbf{B}_2$ is applied on $Q_3$ and $Q_4$.
        }
    \end{center}
\end{figure*}

\begin{figure*}[htb]
\begin{center}
    \includegraphics[width=0.8\textwidth]{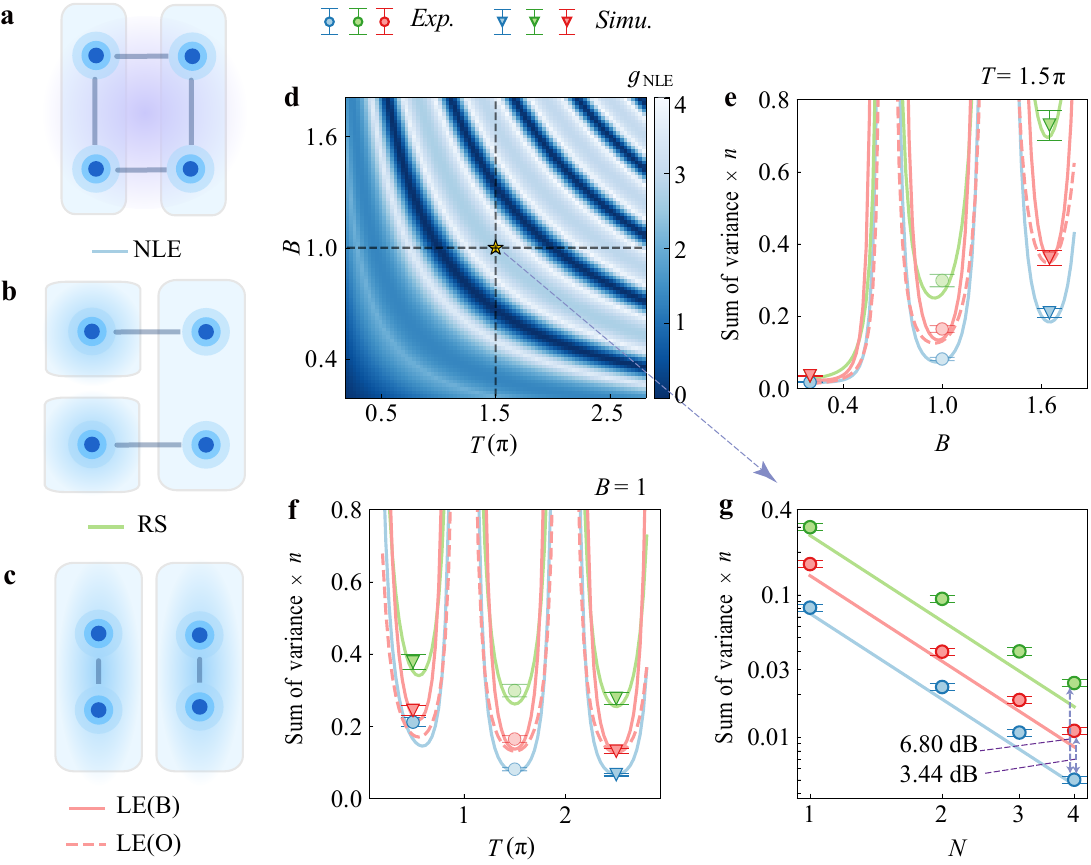}
    \caption{
        \label{fig4SI}
        {\bf Strategies comparison for gradient estimation of a 2-component vector field.}
    {\bf a-c,} Schematic diagrams of different strategies:
    ({\bf a}) Distributed sensing with non-local entanglement (NLE);
    ({\bf b}) Remote sensing (RS) with an ancilla qubit;
    ({\bf c}) Sensing with local entanglement (LE). 
    {\bf d,} Parameter range where NLE outperforms RS and LE. The minimum precision gain of NLE over RS and LE(B) across different $B$ and $T$ values, is calculated by their theoretical precision. 
    {\bf e-f,} Comparison of the precision ($\sum_{i\in \{x,y\}} \delta^2 \nabla\! B_{i_\text{est}}$) of the three strategies.
    ({\bf e}) Precision versus $B$ for the three strategies at $T\ =\ 1.5\pi$ and $N=1$.
    ({\bf f}) Impact of $T$ on estimation precision at $B = 1$ and $N = 1$.
    ({\bf g}) Estimation precision versus $N$ for the three strategies at $T = 1.5\pi$ and $B = 1$.
    The solid and dashed curves: the theoretical precision bound. LE(B): local entanglement strategy using Bell state as the probe state, and Bell measurement. LE(O): local entanglement strategy using the optimal probe state and measurement. The definition of the error bars in (e)-(g) is described in Methods.
    }
\end{center}
\end{figure*}

In addition to the two strategies for gradient sensing discussed above-the distributed strategy with non-local entanglement (NLE) and the local strategy with local entanglement (LE)—we also consider an alternative strategy based on the remote sensing (RS) described in the first section. For this strategy, the remote magnetic fields at each location are estimated independently, with the central module serving as an ancilla, and the gradient is calculated from the separate field estimates. The experimental circuits for these three strategies are illustrated in \rSupFig{separateStrategies}.
For a more comprehensive comparison, we also present the precision of gradient estimation achievable with this alternative strategy. The comparison is made under the condition that the resource allocation is identical, with each sensor module containing two sensors. In \rSupTab{table:precision_2para}, we list the precision achieved by various strategies for estimating gradients along $x$ and $y$ directions.
\red{\rSupFig{fig4SI} further compares their performance by quantifying the estimation precision under various conditions. Specifically, \rSupFig{fig4SI}\textbf{d} identifies the parameter regime where NLE outperforms RS and LE, while \rSupFigs{fig4SI}\textbf{e--g} present systematic comparisons of precision as functions of $B$, $T$, and the cycle numbers $N$.}

\begin{table}[htb]
    \centering
    \renewcommand{\arraystretch}{2.4}
    \setlength{\tabcolsep}{10pt}
    \begin{adjustbox}{width=0.95\textwidth, keepaspectratio}
    \begin{tabular}{lllll}
        \hline\hline
        \noalign{\vskip -2pt}
        \textbf{Strategy} & \textbf{Initial state} & \textbf{Precision} & \textbf{Optimality} & \textbf{Achievability} \\
        \hline
        \textbf{NLE} & 
        $\displaystyle \frac{1}{\sqrt{2}} (|0011\rangle - |1100\rangle)$ & 
        $\displaystyle \frac{1}{4T^2} + \frac{B^2}{4(1 + 3\sin^2(BT))\sin^2(BT)}$ & 
        N & Y \\
        
        \textbf{RS} & 
        $\displaystyle \frac{1}{\sqrt{2}} (|00\rangle + |11\rangle)$ & 
        $\displaystyle \frac{1}{4T^2} + \frac{B^2}{4\sin^2(BT)}$ & 
        Y & Y \\
        
        \multirow{2}{*}{\textbf{LE}} & 
        $\displaystyle \frac{1}{\sqrt{2}}(U_r \otimes U_r)(|11\rangle - |00\rangle)$ & 
        $\displaystyle \frac{1}{8T^2} + \frac{B^2}{8\sin^2(BT)}$ & 
        Y & Y \\
        & $\displaystyle \frac{1}{\sqrt{2}} (|00\rangle + |11\rangle)$ & 
        $\displaystyle \frac{1}{8 \sin^2(BT)} \left( \frac{B^2 - \cos^2(BT) B_x^2}{T^2 B_x^2} + \frac{B^2}{\sin^2(BT)} \right)$ & 
        N & Y \\
        \noalign{\vskip 4pt}
        \hline\hline
    \end{tabular}
    \end{adjustbox}
    \caption{The precision for estimating gradients along $x$ and $y$ directions.}
    \label{table:precision_2para}
\end{table}

\clearpage
\section{Experimental implementation}

\subsection{Device information}
\begin{figure*}[hbt]
    \begin{center}
        \includegraphics[width=1.0\textwidth]{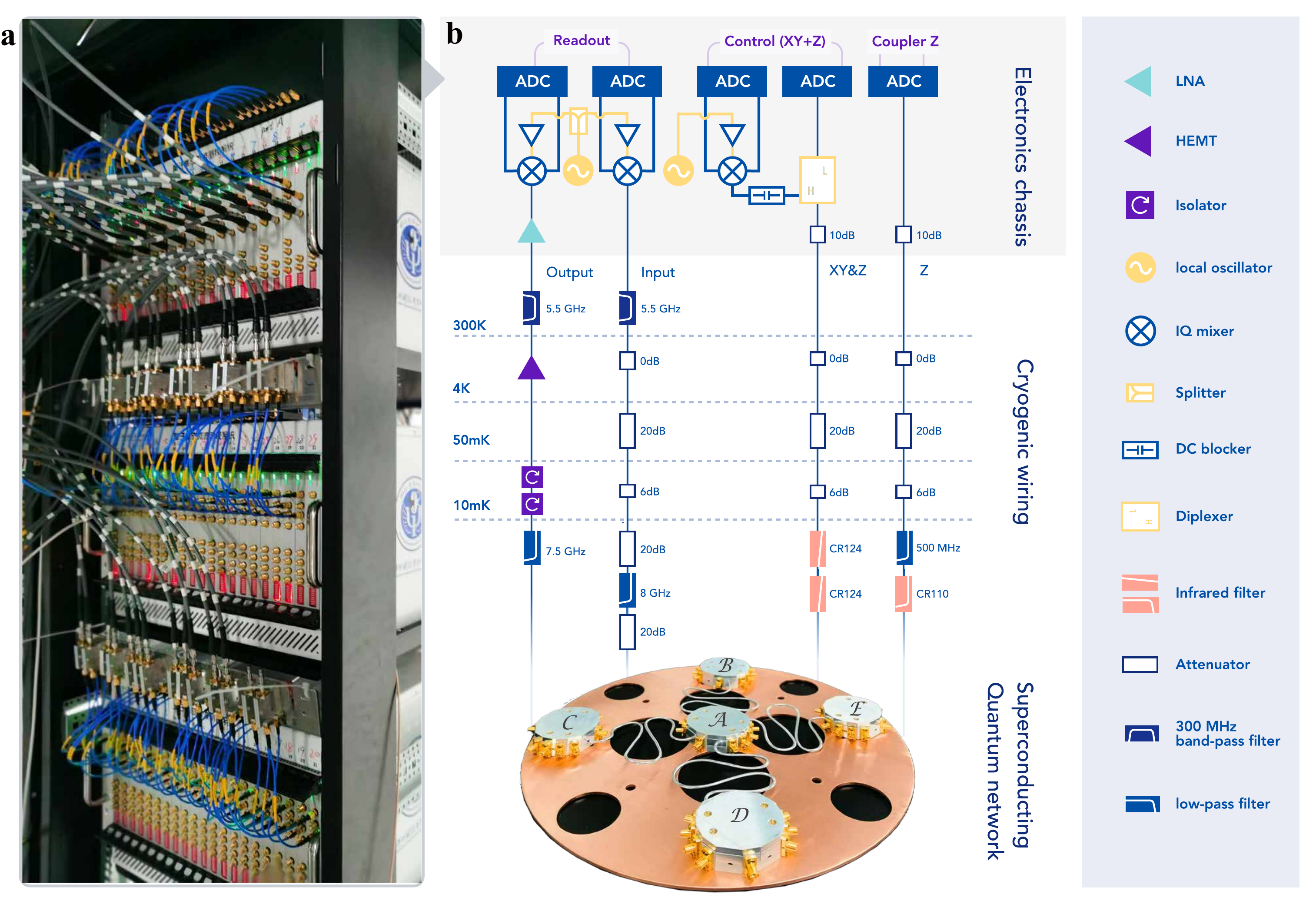}
        \caption{
            \label{setup_wire}
            {\bf The experimental setup.}
            {\bf a:} The microwave control and measurement system built for this experiment.
            {\bf b:} The schematic diagram of the room-temperature electronics chassis, cryogenic wiring, and superconducting quantum network inside the dilution refrigerator. 
            Right panel: the legend of the devices.
        }
    \end{center}
\end{figure*}
We implement the distributed quantum metrology experiment utilizing a modular quantum computing platform composed of five superconducting quantum chips, each integrated with four qubits. The inter-chip connectivity is facilitated by high-quality aluminum superconducting coaxial cables, where a gmon coupler is positioned between the qubits engaged in communication and the cable, enabling tunable coupling strength. Furthermore, an impedance transformer is designed on chip to significantly mitigate the stray loss on the communication channels~\cite{Niu2023}. In \rSupFig{setup_wire}, we show the comprehensive structure of the experimental setup. The distributed quantum processors are sheltered in the 10 mK environment, nestled beneath the mixing chamber of a dilution refrigerator. The microwave cables connecting the superconducting quantum chips serve as the conduit for signal transmission and reception between the quantum processors (see the bottom part of the middle panel) and the customized integrated electronic channels (see the left photograph). The electronics is primarily composed of digital-to-analog converters (DAC) and analog-to-digital converters (ADC), which orchestrate the generation, manipulation and readout of quantum control signals. The generation of XY control signals (single-qubit rotation) is facilitated by IQ mixing of the MHz output of the DACs and the GHz microwave carrier from a local oscillator (LO). Concurrently, the Z control signals (qubit frequency modulation) originate from DC and pulse signals output of the DACs. The XY signals and Z signals belonging to each qubit are combined with a customized diplexer in room temperature. The readout pulses are generated by another set of DACs, LO and IQ mixers, these devices up-convert the probe photons to match the readout resonator frequencies, conversely, the emitted photonic signals from the readout resonators are amplified and down-converted, finally being sampled by the ADCs, completing the readout cycle and providing a digital record of the measurement data. For higher control and readout quality, we deploy multiple filters across different temperature zones within the experimental setup (see the middle panel), the legends detailing the components of these stages are shown in the adjacent panel on the right.

\begin{table}[h]
\centering
\begin{tabular}{c|c|c|c|c|c|c}
\hline
\textbf{Node} & \multicolumn{2}{c}{$\mathcal{A}$} & \multicolumn{2}{|c|}{$\mathcal{B}$} & \multicolumn{2}{c}{$\mathcal{C}$} \\ \hline
\textbf{Qubit} & \textbf{$Q_1$} & \textbf{$Q_2$} & \textbf{$Q_3$} & \textbf{$Q_4$} & \textbf{$Q_5$} & \textbf{$Q_6$} \\ \hline
\textbf{$\omega_\rm{idle}/2\pi\ \rm{(GHz)}$} &\ 4.551\ &\ 5.019\ &\ 4.477\ &\ 4.959\ &\ 4.937\ &\ 4.393\ \\ \hline
\textbf{$\omega_\rm{read}/2\pi\ \rm{(GHz)}$} & 5.629 & 5.692 & 5.686 & 5.627 & 5.688 & 5.621 \\ \hline
\textbf{$E_C/2\pi\ \rm{(MHz)}$} & $-212$ & $-200$ & $-210$ & $-225$ & $-210$ & $-229$ \\ \hline
\textbf{$F_{00}$} & 0.94 & 0.94 & 0.88 & 0.86 & 0.92 & 0.90 \\ \hline
\textbf{$F_{11}$} & 0.91 & 0.90 & 0.87 & 0.83 & 0.88 & 0.89\\ \hline
\textbf{$T_1\ \rm{(\mu s)}$} & 19.2 & 18.4 & 26.8 & 14.3 & 25.3 & 26.2 \\ \hline
\textbf{$T_{2R}\ \rm{(\mu s)}$} & 1.52 & 4.77 & 2.42 & 4.72 & 3.58 & 4.02 \\ \hline
\textbf{$T_{2E}\ \rm{(\mu s)}$} & 5.49 & 10.95 & 11.49 & 15.18 & 15.57 & 14.45 \\ \hline
\textbf{SQG RB fid($\%$)} & 99.96 & 99.91 & 99.95 & 99.47 & 99.80 & 99.75 \\ \hline
\textbf{CZ XEB fid($\%$)} & \multicolumn{2}{c}{98.50} & \multicolumn{2}{|c|}{97.30} & \multicolumn{2}{c}{98.40} \\ \hline
\end{tabular}
\caption{Device information}
\label{qubit_info}
\end{table}

The experiment in this work involves three distributed quantum processors, each containing two qubits. We list the basic information of these six qubits in \rSupTab{qubit_info}. All qubits are designed to be operated across a frequency range of $4.1$ $\sim$ $5.1$~GHz, and idled at staggered frequencies. The resonator frequencies are also staggered for independent readout. The anharmonicity $E_C$ is a parameter determined by the capacitance of each qubit, it is instrumental in shaping the interaction essential for the construction of Controlled-$Z$ (CZ) gates. $F_{00}$ and $F_{11}$ are the state preparation and measurement (SPAM) fidelity for $|0\rangle$ and $|1\rangle$. $T_1$ parameter denotes the energy relaxation time of each qubit, $T_{2R}$ and $T_{2E}$ represent the dephasing time characterized by Ramsey experiment and spin echo experiment, respectively.

\subsection{Gate performance}

{In our experiment, we calibrate and benchmark both local two-qubit gates (CZ, CNOT) and inter-module state transfer operations, together with single-qubit gate performance, in order to establish the reliable entanglement generation and well-characterized metrological gate set.}

The CZ gates were optimized using a standard calibration protocol~\cite{Sung2021}, with a pulse consisting of a 36~ns plateau and 10~ns flattop rising and falling edges. We benchmark the gate performance using cross-entropy benchmarking (XEB)~\cite{Goss2022}, obtaining an average CZ gate fidelity of $98.1\%$.

To enable quantum state transfer across modules, we employ a vacuum Rabi protocol between the communication qubit and the inter-chip cable. This synchronizes the qubit frequencies on both sides of the cable by tuning them into resonance. Simultaneously, the coupler pulses are optimized to achieve the desired effective coupling strength while minimizing reflection losses~\cite{Niu2023}. The fidelity of the transferred state, reconstructed by quantum state tomography (QST), reaches an average of $\sim 98.4\%$. 

The control-signal sequence is implemented by gate sets in the U(3) formalism: $U_3(\alpha,\beta,\lambda)=R_z(\beta)R_x(\pi/2)R_z(\alpha)R_x(-\pi/2)R_z(\lambda)$. The gate set consists of three virtual $Z$ rotations ($\lambda, \alpha, \beta$) interleaved with an $X/2$ gate and a $-X/2$ gate. The single qubit gates (SQG) are benchmarked by randomized benchmarking (RB) experiment, yielding an averaged fidelity of $99.81\%$.

\subsection{Gate set calibration}
\begin{figure*}[hbt]
	\begin{center}
		\includegraphics[width=0.9\textwidth]{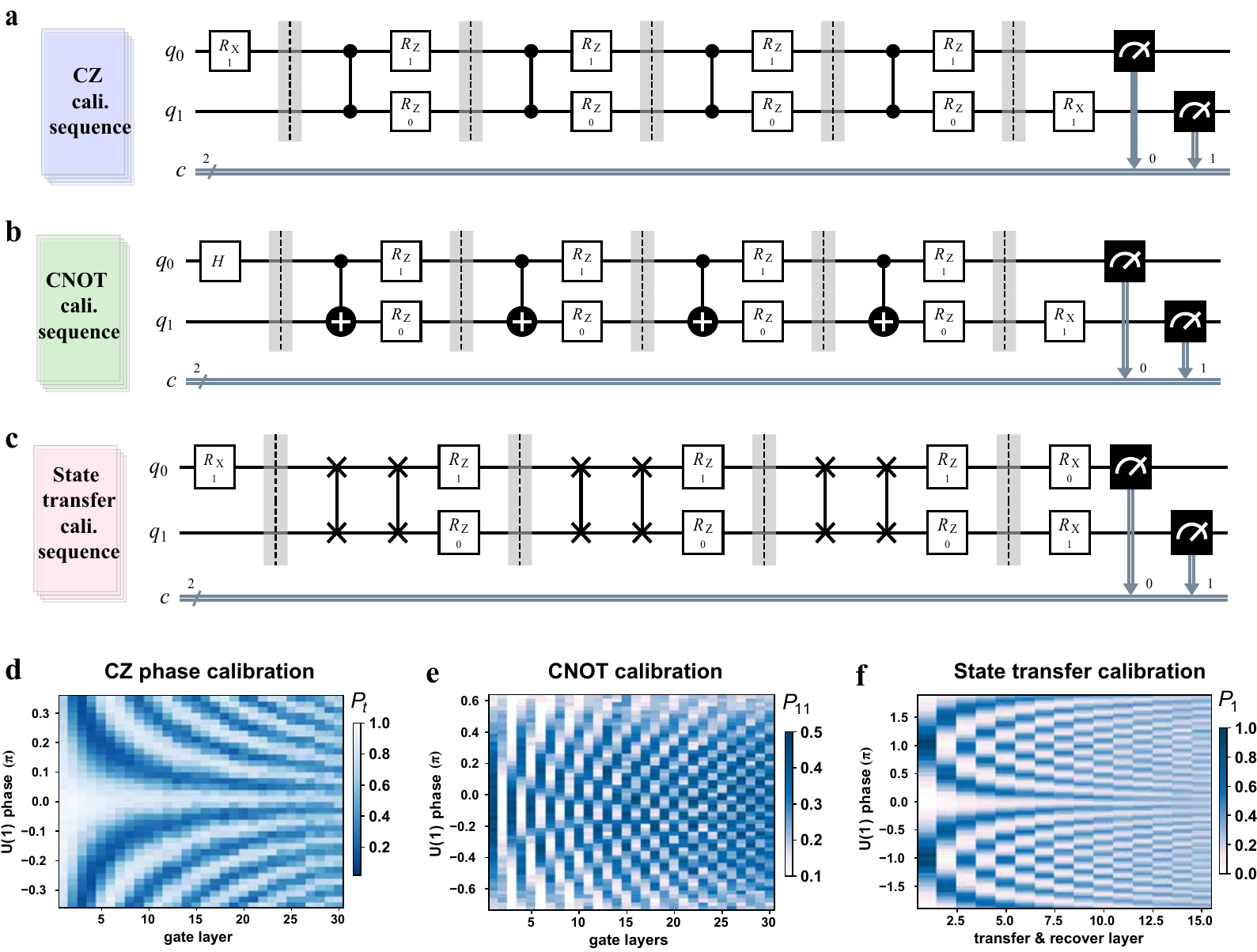}
		\caption{
			\label{Rep_CZcal}
			{\bf {Repeated gate sequences (drawn by Qiskit ~\cite{JavadiAbhari2024}) for $U(1)$ phase calibration.} }
			{\bf a,} CZ phase calibration sequence. The virtual $Z$ phase of the control qubit is fixed, while that of the target qubit is swept, and the resulting probability on the target qubit ($P_t$) is measured to extract the accumulated phase. 
           {\bf b,} CNOT phase calibration sequence. A CNOT gate is compiled into a $-Y/2$ gate, followed by a CZ gate and a $Y/2$ gate. The control qubit is initialized in the $|+\rangle$ state by a Hadamard gate, and repeated CNOT operations reveal oscillations between $\tfrac{1}{\sqrt{2}}(|00\rangle+|11\rangle)$ and $|00\rangle$, with the correct compensating phase maximizing the exchange, exhibited by the probability on state $|11\rangle$ ($P_{\rm{11}}$).
           {\bf c,} Calibration sequence for inter-module quantum state transfer. We transfer a $|0\rangle + i|1\rangle$ state from the transmission qubit ($q_0$) to the receiving qubit ($q_1$), and then use another state transfer to recover it back to the transmission qubit. This transfer and recover set is repeated by multiple layers, and the result is measured on X-axis of $q_1$.
            {\bf d–f,} Extraction of virtual $Z$ phases from repeated sequences of CZ, CNOT gates, and state transfer operations, respectively.
		}
	\end{center}
\end{figure*}

{To further ensure accurate multi-qubit control, we carefully calibrate the phases associated with entangling gates. 
An ideal CZ gate applies only a $\pi$ phase to the $|11\rangle$ state, while leaving other computational basis states unchanged. In practice, however, frequency detuning between the two qubits leads to additional single-qubit phases on $|01\rangle$ and $|10\rangle$, which manifest as relative $U(1)$ rotations in the two-qubit rotating frame.
We characterize these phases using Ramsey sequences, and compensate them with appropriate virtual-$Z$ corrections. As shown in \rSupFig{Rep_CZcal}{\bf a}, by repeating layers of phase-compensated CZ gates, we observe oscillatory interference fringes between $|0\rangle$ and $|1\rangle$ states, allowing us to identify the phase setting that ensures constructive alignment as the number of two-qubit gate layers increases (\rSupFig{Rep_CZcal}{\bf d}).  }

{Similar phase accumulation effects arise not only in CZ gates but also in CNOT and remote state transfer operations. We quantify and compensate these phases using Ramsey-type sequences (\rSupFig{Rep_CZcal} {\bf a-c}), which reveal increasing sensitivity with the number of repeated operations. By sweeping the compensation phase, we obtain population profiles that converge to stable values (\rSupFig{Rep_CZcal} {\bf d-f}), from which the optimal corrections can be extracted. These corrections are then consistently applied across the full gate set, ensuring that all local and non-local operations remained phase-aligned. 
}

\subsection{Implementation of distributed sensing in quantum circuits}

\begin{figure*}[hbt]
	\begin{center}
		\includegraphics[width=1.0\textwidth]{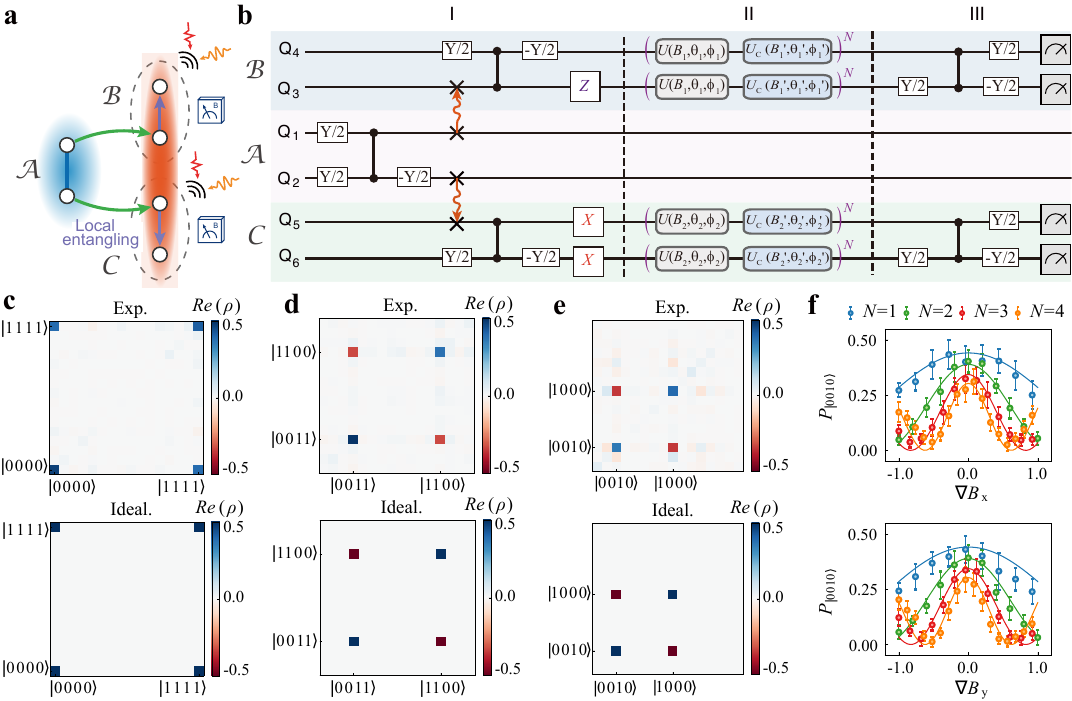}
		\caption{
        \label{circuit_6Q}
            {\bf a,} A brief schematic diagram of NLE metrology strategy for gradient.
            {\bf b,} The detailed quantum circuit of (a).
            {\bf c,} The remote four-qubit GHZ state generated across three modules.
            {\bf d,} The four-qubit probe state $|\Psi_{0}\rangle=\frac{1}{\sqrt{2}}(|0011\rangle-|1100\rangle)$.
            {\bf e,} The reference final state $|\Psi_{f}\rangle = \frac{1}{\sqrt{2}}(|0010\rangle - |1000\rangle)$ after step III, with no control and signal units inserted into the circuit.
            {\bf f,} The probability oscillation observed when scanning parameter $\nabla B_{x}$ or $\nabla B_{y}$ for different $N$, the encoding time is fixed at $T=0.2\pi$. The error bars denote the standard deviation.
		}
	\end{center}
\end{figure*}

\begin{figure*}[hbt]
	\begin{center}
		\includegraphics[width=1.0\textwidth]{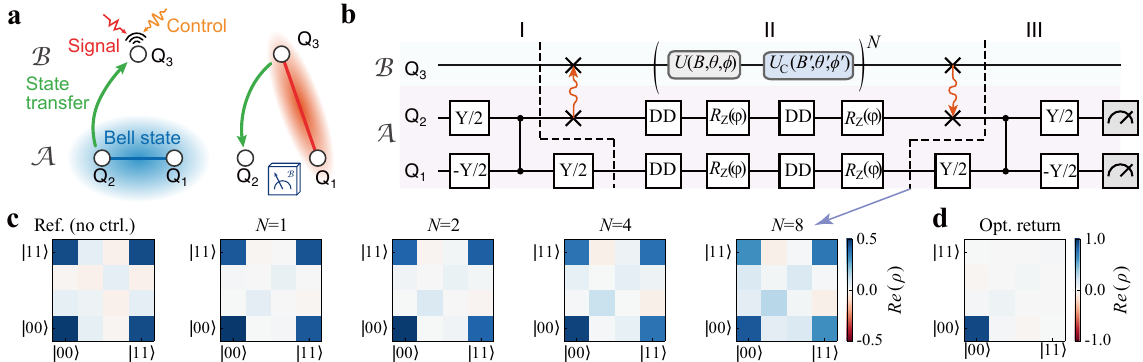}
		\caption{
			\label{circuit_3Q}
			{\bf a,} A brief schematic diagram for sensing a local vector field at a certain position with two nodes $\mathcal{A}$ and $\mathcal{B}$. 
            {\bf b,} The quantum circuit diagram for {\bf a}. I: state preparation, II: encoding, III: measurement.
            {\bf c,} Ref (no ctrl): the density matrix with no signal-control sequence, measured after step II. $N=1\sim 4$: the density matrices with $N=1,2,4,8$, measured after step II.
            {\bf d,} Opt return: the density matrix with no signal-control sequence, measured after step III.
		}
	\end{center}
\end{figure*}

The NLE strategy for gradient metrology is implemented using the three-node sensor network illustrated in \rSupFig{circuit_6Q}, composed of modules $\mathcal{A}$, $\mathcal{B}$, and $\mathcal{C}$ (\rSupFig{circuit_6Q}{\bf a}). As illustrated in \rSupFig{circuit_6Q}{\bf b}, the protocol begins by generating a Bell pair between $Q_1$ and $Q_2$ on module $\mathcal{A}$. Subsequently, the two qubits of this pair are transferred simultaneously: one from $Q_1$ to $Q_3$ on module $\mathcal{B}$, and the other from $Q_2$ to $Q_5$ on module $\mathcal{C}$. CNOT gates are then applied on both modules $\mathcal{B}$ and $\mathcal{C}$, resulting in a GHZ state across the two modules with a fidelity of $80.36\%$ (\rSupFig{circuit_6Q}{\bf c}). The probe state $|\Psi_{0}\rangle$ is prepared by applying additional X gates on $Q_3$ and $Q_4$, and a Z gate on $Q_5$, achieving a fidelity of $76.16\%$ (see \rSupFig{circuit_6Q}{\bf d}). Subsequently, we encode the spatially distributed vector field on sensor chips $\mathcal{B}$ and $\mathcal{C}$ with $U(3)$-formalism gate sets, where $\mathbf{B}_1$ acts on both $Q_3$ and $Q_4$, $\mathbf{B}_2$ acts on both $Q_5$ and $Q_6$. Following the encoding process, we conduct Bell measurement on both modules $\mathcal{B}$ and $\mathcal{C}$. Under optimal control, the ideal final state $|\Psi_f\rangle$ has equivalent occupation on $|0010\rangle$ and $|1000\rangle$, the fidelity obtained from the experiment is $75.20\%$ (see \rSupFig{circuit_6Q}{\bf e}). 
The information carried by the sensors is decoded into the probability distribution in the measurement basis (see \rSupFig{circuit_6Q}{\bf f} for $P_{0010}$). The oscillation period of this probability with respect to the gradient components $\nabla B_{x}$ and $\nabla B_{y}$ is observed to contract as the number of sequential copies $N$ increases. 

The error in this sequence primarily stems from the control error when synchronously transferring two entangled states. The control error is estimated to be $11.44\%$ for generating the non-local GHZ state; the decoherence error throughout this 340 ns sequence is approximately $8.34\%$. These two parts yield an estimated fidelity of $80.22\%$, which is close to our experimental result of $80.36\%$. The non-local entangled state across two chips, which are not directly connected, is more fragile to environmental noise. The effective decoherence rate of the probe state is estimated to be $80\times 2\pi\ \mathrm{kHz}$. As a consequence, the fidelity values of $|\Psi_f\rangle$ are $69.27\%, 63.81\%, 58.78\%, 54.15\%$ for $N=1\sim 4$, respectively. These values are higher than the confidence threshold $\sim50\%$ for entanglement. 

Demonstrating the remote sensing strategy requires pairs of distributed nodes with high-quality connections. Using the $\mathcal{A}$-$\mathcal{B}$ pair as an example (Fig. 1{\bf a} in the main text), a locally generated entangled state (blue shadow) is transferred to the remote node $\mathcal{B}$, establishing cross-module entanglement (red shadow). This setup forms the basis of our conceived metrology network: a central module ($\mathcal{A}$) is connected to multiple remote sensor modules ($\mathcal{B}, \mathcal{C}$, etc.), which are spatially positioned to perform distributed sensing of a local field.

The quantum circuit is illustrated in \rSupFig{circuit_3Q}{\bf b}. We first generate a local Bell state on chip $\mathcal{A}$, and subsequently transfer the state of one qubit, $Q_2$ on chip $\mathcal{A}$ to $Q_3$ on module $\mathcal{B}$, resulting in a cross-node Bell state between $Q_1$ and $Q_3$. $Q_3$ is then used as the sensor qubit and $Q_1$ as the ancillary qubit. To protect the inter-module entanglement from fast decoherence during the sensing interval, we apply dynamical decoupling sequences to $Q_1$ and $Q_2$. The quantum state is subsequently transferred from the sensor qubit ($Q_3$) back to $Q_2$, followed by a Bell measurement on $Q_1$, $Q_2$. Under optimal control, the entire Step II functions as an identity operation. We perform quantum state tomography on module $\mathcal{A}$ after state retrieval. \rSupFig{circuit_3Q}{\bf c} displays the extracted real parts of the density matrices for $Q_2$ and $Q_3$.  

To capture errors from state preparation and measurement (SPAM), we implement a reference circuit. The reference circuit is obtained from the original ciruit by excluding signal encoding, control operations, dynamical decoupling, and phase compensation (\(R_z\) gates), i.e., only keeps the state preparation and measurement. The reference circuit achieves a fidelity of \(91.15\%\) for the preparation of the probe state, and after including additional gates required for Bell-basis measurement, it yields a fidelity of \(90.52\%\). In contrast, circuits incorporating signal encoding and optimal control sequences—with dynamical decoupling and \(R_z\) gates for phase compensation—were evaluated for \(N = 1, 2, 4, 8\) repetitions of the signal-control unit. The corresponding state fidelities are \(88.13\%\), \(84.37\%\), \(79.02\%\), and \(70.46\%\), respectively. The decline in fidelity with increasing \(N\) reflects error accumulation during the sensing protocol. By comparing these results with the reference fidelity, we can separate SPAM errors from those introduced by the signal-control sequence. The total error per signal-control unit is estimated at approximately \(3.34\%\), comprising an average control error of \(0.79\%\) per \(U_x\) or \(U_c\) operation and a decoherence contribution of \(2.55\%\), the latter corresponding to an effective decoherence rate of \(41.1 \times 2\pi~\mathrm{kHz}\).

\begin{figure*}[hbt]
	\begin{center}
		\includegraphics[width=0.8\textwidth]{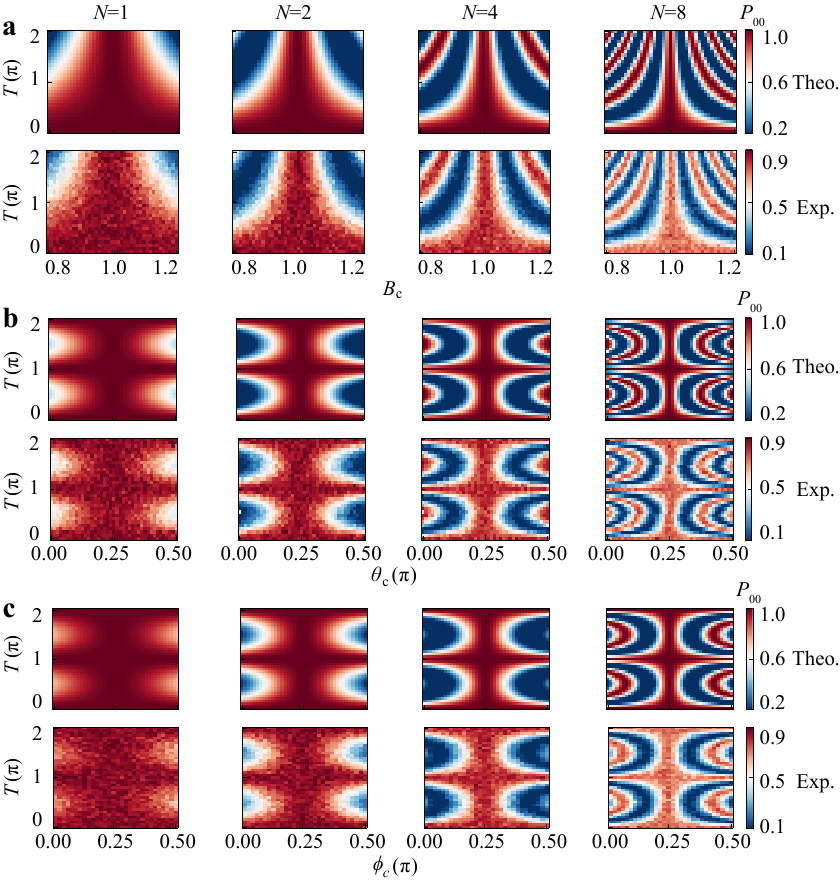}
		\caption{
			\label{para_prob_T}
			{\bf The probability oscillation with three parameters and encoding time $T$.}
        {\bf a,} For a fixed signal, we scan control parameter $B_c$ and encoding time $T$ for different $N$. 
        {\bf b,} We scan control parameter $\theta_c$ and encoding time $T$ for different $N$.
        {\bf c,} We scan control parameter $\phi_c$ and encoding time $T$ for different $N$.
		}
	\end{center}
\end{figure*}

A key feature of the relationship between the number of control-signal layers ($N$) and the estimated precision is evident in the oscillation period of the probability derived from Bell measurement. As shown in \rSupFig{para_prob_T}, the period of the $P_{00}$ profile near the optimal control parameters is inversely proportional to $N$. Scanning a single parameter is analogous to a single-parameter estimation process. In this context, the quantum Fisher information is determined by the derivative of the probability with respect to the target parameter. The reduction in oscillation period with increasing $N$ thus provides an intuitive understanding of the enhancement offered by the sequential strategy. 
Moreover, the encoding time $T$ is also periodically correlated to the probability distribution, and this correlation depends on the signal parameters. For example, when the signal is set to $(B,\theta,\phi)=(1,\pi/4,\pi/4)$, the period of $P_{00}$ profile at $T=\pi,2\pi$ stays invariant with respect to the parameters $\theta$ and $\phi$, regardless of $N$. In contrast, at $T=0.5\pi$ or $T=1.5\pi$, the oscillation period changes more dramatically, indicating that the sequential strategy is effective there.

To implement LE strategy, we generate local Bell state on node $\mathcal{B}$ and $\mathcal{C}$, signals and controls are simultaneously acting on two pairs of sensor qubits. The sequence ends with Bell measurement on $\mathcal{B}$ and $\mathcal{C}$, which is similar to RS strategy.


\clearpage

\section{Extended data}

\subsection{Extended data for sensing of remote vector fields}

\begin{figure*}[hbt]
	\begin{center}
		\includegraphics[width=0.8\textwidth]{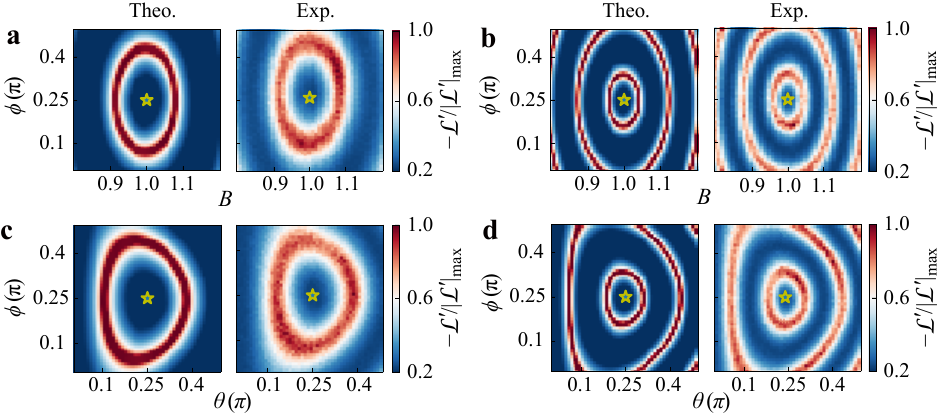}
		\caption{
			\label{2m3c_landscape}
			{\bf The likelihood function landscape at $N=4$ and $N=8$.} Stars: the location of the optimal control parameters.
        {\bf a,} The landscape for parameter $B$ and $\phi$ at $N=4$.
        {\bf b,} The landscape for parameter $B$ and $\phi$ at $N=8$.
        {\bf c,} The landscape for parameter $\theta$ and $\phi$ at $N=4$.
        {\bf d,} The landscape for parameter $\theta$ and $\phi$ at $N=8$.
        We post the theoretical landscape and experimental result, and mark the optimal control parameters.
		}
	\end{center}
\end{figure*}
 
We benchmark the sensor-ancilla network by analyzing the landscape of the likelihood function $\mathcal{L}^{\prime}$ near the optimal control parameters, the results are shown over two variables in \rSupFig{2m3c_landscape}. Specifically, the panels depict $\mathcal{L}^{\prime}(B,\phi)$ in \rSupFigs{2m3c_landscape}{\bf a,b} and $\mathcal{L}^{\prime}(\theta,\phi)$ in \rSupFigs{2m3c_landscape}{\bf c,d}. 
The optimal control parameters, marked with a star in each panel, correspond to the expected estimation results. As $N$ increases, the boundary area of the likelihood landscape contracts, demonstrated for $N=4$ in \rSupFigs{2m3c_landscape}{\bf a,c} and $N=8$ in \rSupFigs{2m3c_landscape}{\bf b,d}. The agreement between theoretical and experimental results ensures that the estimated parameters not only align with the observed data but also adhere to the underlying physical model. {Moreover, in a qualitative perspective, the slight blurring and distortion on the contour of the experimental landscape, comparing to the ideal results, reflect the impact of decoherence and control errors on the protocol, respectively.}
\begin{figure*}[hbt]
	\begin{center}
		\includegraphics[width=0.7\textwidth]{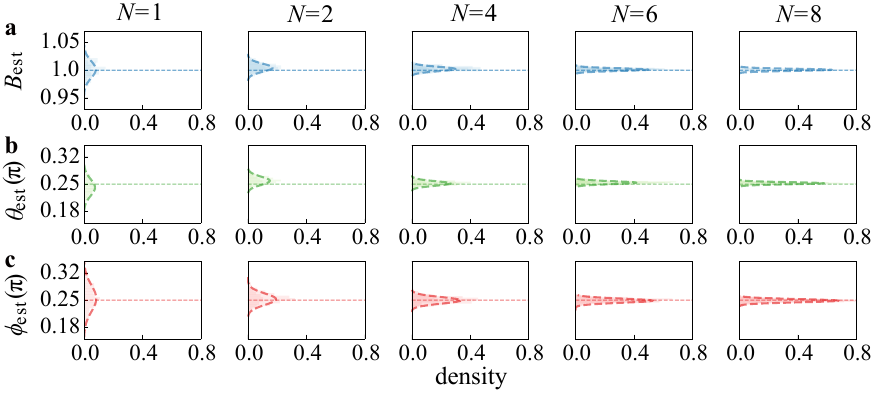}
		\caption{
			\label{2m3c_density}
			{\bf The result for simultaneously estimating three parameters of a local vector field $\mathbf{B}(B,\theta,\phi)$.} Bars: MLE result histograms. Dashed curves: gaussian fitting of the histograms. Dashed lines: the ideal signal parameters.
        {\bf a,} The density distribution of parameter $B$.
        {\bf b,} The density distribution of parameter $\theta$.
        {\bf c,} The density distribution of parameter $\phi$.
		}
	\end{center}
\end{figure*}

\begin{figure*}[hbt]
	\begin{center}
		\includegraphics[width=0.8\textwidth]{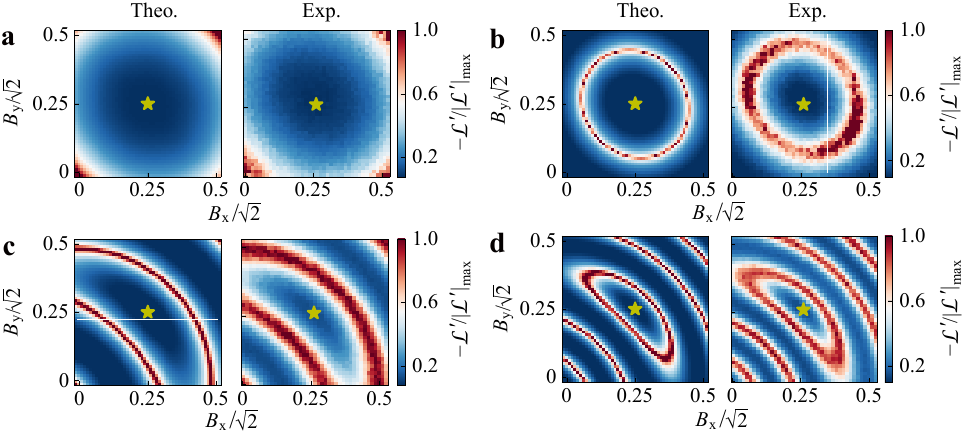}
		\caption{
			\label{2m2c_B=0.5_landscape}
			{\bf The theoretical and experimental landscape of estimating a two-component vector field with a sensor-ancilla network.} Stars: the location of the optimal control parameters.
        {\bf a,} At $N=2$ and $T=0.5\pi$.
        {\bf b,} At $N=4$ and $T=0.5\pi$.
        {\bf c,} At $N=2$ and $T=1.5\pi$.
        {\bf d,} At $N=4$ and $T=1.5\pi$
		}
	\end{center}
\end{figure*}

The \rSupFig{2m3c_density} illustrates the complete MLE result for sensing the three-component vector field. The density amplitude in each panel represents the count of the distribution normalized by the integral of the distribution. These results demonstrate that increasing the number of sequential copies enhances the precision of simultaneous three-parameter estimation.
However, unavoidable experimental errors introduce a bias of up to $\pm 3.45\%$ in the averaged estimation results, deviating from the actual vector field parameters. 
These errors can distort the likelihood landscape, affecting the efficiency of the MLE process.
Moreover, fluctuation in experimental noise lead to inhomogeneous probability distributions, increasing the risk of convergence to local rather than global minima.
Despite these challenges, the optimal strategy we implement facilitates a flat region around the optimal control parameters, and the use of multiple initial values mitigates the impact of local minima~\cite{Kuroda2020}.
This approach improves the robustness and reliability of the estimation process~\cite{Hou2021}.

We apply the same benchmarking approach to a two-component vector field.
As shown in \rSupFig{2m2c_B=0.5_landscape}, the contraction of the likelihood function landscape with increasing $N$ is evident. The parameters to be estimated in this case are set as $\mathbf{B} = (\frac{\sqrt{2}}{2}, \frac{\sqrt{2}}{2}, 0)$, corresponding to $|\mathbf{B}|=0.5$.

\subsection{Extended data for distributed sensing of vector field gradient}

\begin{figure*}[hbt]
	\centering
	\includegraphics[width=0.75\textwidth]{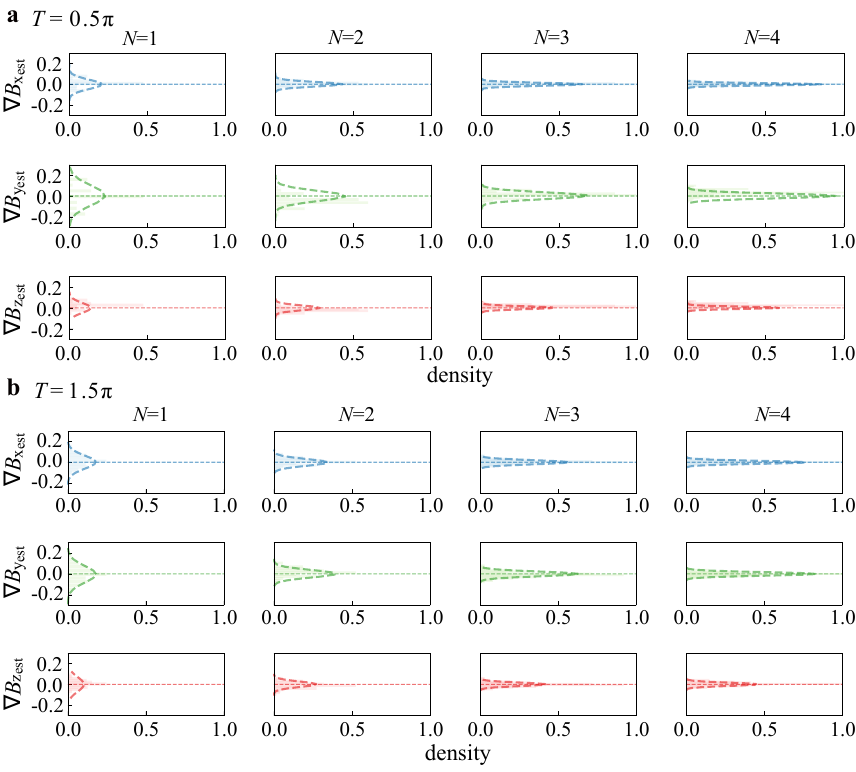}
	\caption{
        {\bf The normalized distribution of estimators for simultaneously three-component estimation.} Bars: MLE result histograms. Dashed curves: gaussian fitting of the histograms. Dashed lines: the ideal signal parameters.
        {\bf a,} At $T=0.5\pi$ and $N=1\sim 4$.
        {\bf b,} At $T=1.5\pi$ and $N=1\sim 4$.
            }
	\label{3m3c_density}
\end{figure*}
 
We experimentally evaluate the performance of NLE strategy for simultaneously estimating the three components of the gradient. In this scheme, we set $\mathbf{B}=(\frac{1}{2},\frac{1}{2},\frac{\sqrt{2}}{2})$, the normalized distributions of the estimators $\nabla B_{x_\rm{est}}$, $\nabla B_{y_\rm{est}}$, and $\nabla B_{z_\rm{est}}$ are shown in \rSupFig{3m3c_density}{\bf a} (for $T=0.5\pi$) and \rSupFig{3m3c_density}{\bf b} (for $T=1.5\pi$).

\begin{figure*}[htb]
	\centering
	\includegraphics[width=0.65\textwidth]{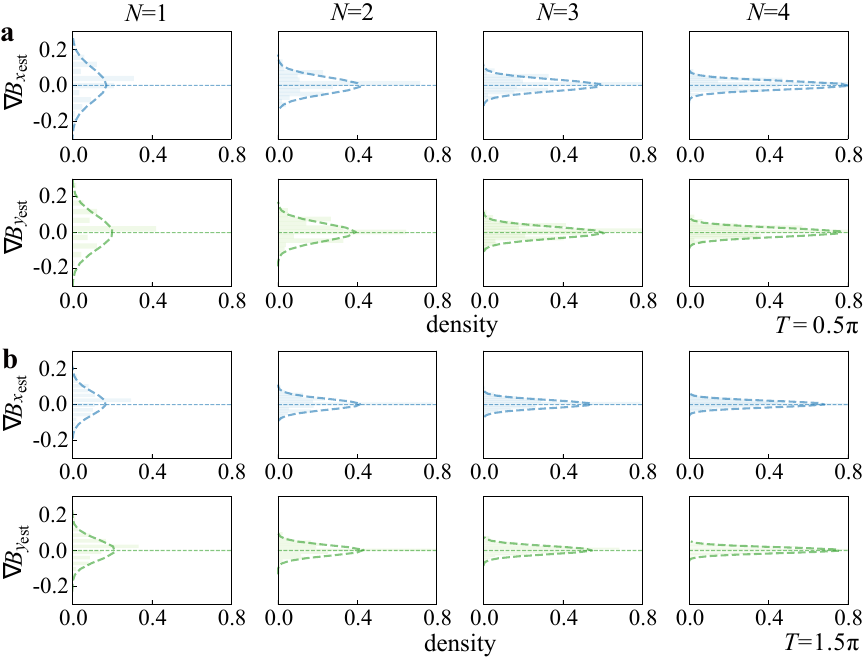}
	\caption{
        {\bf The density distribution for estimated $x$ and $y$ components of the vector field gradient.} Bars: MLE result histograms. Dashed curves: gaussian fitting of the histograms. Dashed lines: the ideal signal parameters.
        {\bf a,} The density at $T=0.5\pi$ for $N=1,2,3,4$.
        {\bf b,} The density at $T=1.5\pi$ for $N=1,2,3,4$.
            }
	\label{3m2c_density}
\end{figure*}

When estimating the gradient of a two-component vector field $\nabla \mathbf{B} = (\nabla B_x, \nabla B_y)$, the field amplitudes are expressed as a function of gradient $\nabla \mathbf{B}$ and sum $\sum\!\mathbf{B}$ at two distinct positions, with $\mathbf{B}_1 = (\sum\!\mathbf{B} + \nabla \mathbf{B})/2$ and $\mathbf{B}_2 = (\sum\!\mathbf{B} - \nabla \mathbf{B})/2$. 
The full dataset, presented in Fig.~2 of the main text, is shown in \rSupFig{3m2c_density}, where the $x$ and $y$ components are plotted separately. 
The signal parameters are chosen as $\sum\!\mathbf{B} = (\sqrt{2}/2, \sqrt{2}/2, 0)$ and $\nabla \mathbf{B} = (0,0,0)$. The encoding times are set to $T=0.5\pi$ (\rSupFig{3m2c_density}{\bf a}) and $T=1.5\pi$ (\rSupFig{3m2c_density}{\bf b}).

\subsection{The influence of noise}

\begin{figure*}[hbt]
	\centering
	\includegraphics[width=0.7\textwidth]{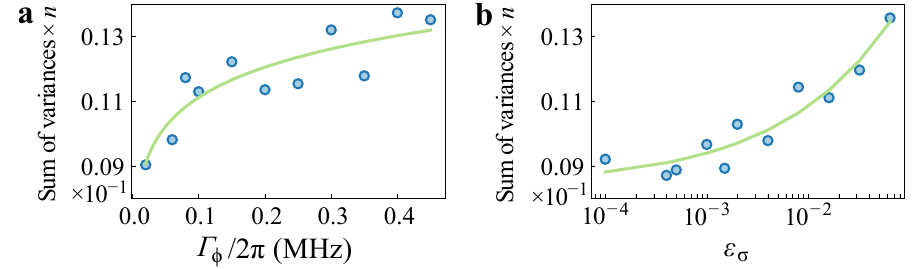}
	\caption{ 
        {\bf The influence of circuit error on the MLE result.} Dots: simulation data. Solid curves: fitting result.
        {\bf a} The relation between simulated precision and dephasing rate.
        {\bf b} The relation between simulated precision and gate error.
            }
	\label{error_simu}
\end{figure*}

Different types of noise in quantum system have impact to the precision of the gradiometer. The effects of noisy channels in quantum parameter estimation have been discussed in previous studies~\cite{Le2023,Yuan2017,Escher2011,Wang2018,Chaves2013,Peng2024}. 
In our work, the dominant sources of noise are control errors and dephasing. We numerically simulate the {quantitative} relationship between these noise types and the precision of the estimation.
\rSupFig{error_simu}{\bf a} shows the sum of variances in the estimated gradient as a function of different dephasing rates $\Gamma_{\phi}$, while \rSupFig{error_simu}{\bf b} depicts the effect of varying gate errors $\epsilon_{\sigma}$ on the precision. The dephasing noise is modeled using a thermal channel, and gate errors are incorporated through a Pauli noise channel. These simulations are conducted using the Qiskit framework~\cite{JavadiAbhari2024}. 
\begin{figure}[hbt]
	\centering
	\includegraphics[width=0.42\textwidth]{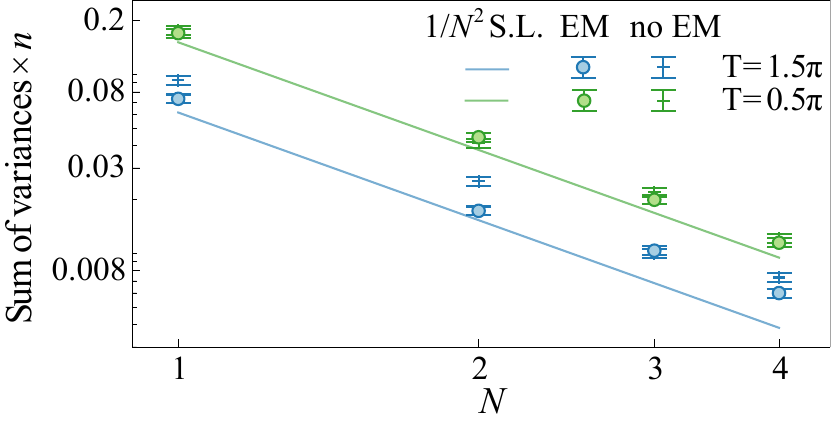}
	\caption{ 
        {\bf The effect of error mitigation (EM) on the experimental sum of variances.} Different markers indicate data obtained with EM and without EM. The definition of the error bars is described in Methods.
            }
	\label{error_mitigation}
\end{figure}

To mitigate the effects of noise, we apply error mitigation (EM) techniques~\cite{Conlon2023} during data post-processing. For the NLE strategy, non-local entangled states are particularly sensitive to environmental noise. However, EM significantly improves performance, as demonstrated in \rSupFig{error_mitigation}.

\clearpage
\renewcommand{\refname}{References}
\renewcommand{\bibname}{References}